\documentclass[aps,pra,reprint,showpacs,superscriptaddress,10pt]{revtex4-1}
\pdfoutput=1
\usepackage{amsmath,amssymb,amsfonts,graphicx,bm,longtable,dcolumn,color,ulem}
\usepackage{mathrsfs}
\usepackage[utf8]{inputenc}
\usepackage{textcomp}
\usepackage[T1]{fontenc}
\usepackage{lmodern}

\allowdisplaybreaks

\begin{document}

\title{Radiative heat transfer and nonequilibrium Casimir-Lifshitz force\\in many-body systems with planar geometry}

\author{Ivan Latella}
\email{ivan.latella@institutoptique.fr}
\affiliation{Laboratoire Charles Fabry, UMR 8501, Institut d'Optique, CNRS, Universit\'{e} Paris-Saclay, 2 Avenue Augustin Fresnel, 91127 Palaiseau Cedex, France}

\author{Philippe Ben-Abdallah}
\email{pba@institutoptique.fr}
\affiliation{Laboratoire Charles Fabry, UMR 8501, Institut d'Optique, CNRS, Universit\'{e} Paris-Saclay, 2 Avenue Augustin Fresnel, 91127 Palaiseau Cedex, France}
\affiliation{Universit\'{e} de Sherbrooke, Department of Mechanical Engineering, Sherbrooke, PQ J1K 2R1, Canada.}

\author{Svend-Age Biehs}
\email{s.age.biehs@uni-oldenburg.de}
\affiliation{Institut f\"{u}r Physik, Carl von Ossietzky Universit\"{a}t, D-26111 Oldenburg, Germany}

\author{Mauro Antezza}
\email{mauro.antezza@umontpellier.fr}
\affiliation{Laboratoire Charles Coulomb (L2C), UMR 5221 CNRS-Universit\'{e} de Montpellier, F- 34095 Montpellier, France}
\affiliation{Institut Universitaire de France, 1 rue Descartes, F-75231 Paris Cedex 05, France}

\author{Riccardo Messina}
\email{riccardo.messina@umontpellier.fr}
\affiliation{Laboratoire Charles Coulomb (L2C), UMR 5221 CNRS-Universit\'{e} de Montpellier, F- 34095 Montpellier, France}

\begin{abstract}
A general theory of photon-mediated energy and momentum transfer in $N$-body planar systems out of thermal equilibrium is introduced. It is based on the combination of the scattering theory and the fluctuational-electrodynamics approach in many-body systems. By making a Landauer-like formulation of the heat transfer problem, explicit formulas for the energy transmission coefficients between two distinct slabs as well as the self-coupling coefficients are derived and expressed in terms of the reflection and transmission coefficients of the single bodies. We also show how to calculate local equilibrium temperatures in such systems. An analogous formulation is introduced to quantify momentum transfer coefficients describing Casimir-Lifshitz forces out of thermal equilibrium. Forces at thermal equilibrium are readily obtained as a particular case. As an illustration of this general theoretical framework, we show on three-body systems how the presence of a fourth slab can impact equilibrium temperatures in heat-transfer problems and equilibrium positions resulting from the forces acting on the system. 
\end{abstract}

\newcommand{\dif}{d}
\newcommand{\iunit}{i}
\newcommand{\sub}[1]{#1}
\newcommand{\vect}[1]{\bm{#1}}
\newcommand{\kB}{k_B}

\maketitle

\section{Introduction}

\vspace{-2mm}
Fluctuations of the electromagnetic field are responsible for momentum~\cite{CasimirProcKNedAkadWet48,CasimirPhysRev48,Lifshitz55} and heat~\cite{PoldervH} exchanges. The pioneering works of Casimir~\cite{CasimirProcKNedAkadWet48} and Casimir and Polder~\cite{CasimirPhysRev48} were the first showing that such fluctuations are at the origin of an attractive force between two perfectly conducting infinite parallel planes as well as between an atom and a plate. This effect was theoretically predicted to exist even at thermal equilibrium, at zero temperature and in vacuum. A few years later, Lifshitz~\cite{Lifshitz55} and subsequently Dzyaloshinskii, Lifshitz, and Pitaevskii~\cite{DzyaloshinskiiAdvPhys61} developed a more general theory taking into account real material properties and thermal effects. Much more recently, a series of papers~\cite{AntrezzaPRA04,AntezzaPRL05,AntezzaPRL06,AntezzaPRA08} were focused on the effect of the absence of thermal equilibrium, showing that the presence of different temperatures in the system can not only qualitatively modify the behavior of the force (e.g. changing its power-law dependence on the distance), but also induce a repulsive force, otherwise impossible for standard geometries such as two parallel slabs. Starting from 1997, several experiments have confirmed the theoretical predictions at thermal equilibrium, both for configurations involving macroscopic bodies (mainly in the plane-plane and sphere-plane geometries)~\cite{LamoreauxPRL97,MohideenPRL98,RoyPRD99,EderthPRA00,ChanPRL01,ChanScience01,BressiPRL02,DeccaPRL03,DeccaAnnPhys05,HarberPRA05,DeccaPRD07,KrausePRL07,CapassoIEEE07,PalasantzasAPL08,ChanPRL08,MundayPRA08,MundayNature09,JourdanEPL09,deManPRL09,ChiuPRB10,SushkovNatPhys11,ZuurbierNJP11,BimontePRB16} and in the atom-plane configuration~\cite{deflection,Aspect1996,Shimizu2001,DeKieviet2003,Zao10,Pasquini1,Pasquini2}. Recently, the force between a BEC and a plane has also been successfully measured out of thermal equilibrium~\cite{ObrechtPRL07}, confirming the theoretical predictions.

Another phenomenon emerging as a consequence of absence of thermal equilibrium is radiative heat transfer. Using Rytov's theory of fluctuational electrodynamics~\cite{Rytov}, Polder and van Hove~\cite{PoldervH} showed that this energy-exchange mechanism, limited in the far field (for distances much larger than the thermal wavelength $\lambda_T=\hbar c/k_BT$, close to 8\,$\mu$m at room temperature) by the blackbody limit predicted by Stefan-Boltzmann's law, can overcome this value even by several orders of magnitude in the near-field regime. In particular, it was later shown that this amplification is particularly pronounced for materials supporting surface modes such as phonon polaritons~\cite{JoulainSurfSciRep05,VolokitinRMP07}. The radiative heat transfer has been experimentally investigated as well~\cite{HargreavesPLA69,KittelPRL05,HuApplPhysLett08,NarayanaswamyPRB08,RousseauNaturePhoton09,ShenNanoLetters09,KralikRevSciInstrum11,OttensPRL11,vanZwolPRL12a,vanZwolPRL12b,KralikPRL12,KimNature15,StGelaisNatureNano16,SongNatureNano15,KloppstecharXiv,WatjenAPL16}, in a range of distances going from the nanometer region to several microns, confirming the behavior predicted theoretically.

The recent theoretical history on both topics has seen the development of a series of general theories for Casimir forces~\cite{TomasPRA02,RaabePRA03,BezerraEPJC07,RahiPRD09,ReidPRL09,RodriguezPRA09,ReidPRA11,ReidPRA13}, radiative heat transfer~\cite{RodriguezPRL11,McCauleyPRB12,RodriguezPRB12,RodriguezPRB13,NarayanaswamyJQSRT14,MullerarXiv} or both in a unified approach~\cite{BimontePRA09,KrugerPRL11,MessinaEurophysLett11,MessinaPRA11,KrugerPRB12,MessinaPRA14,BimontearXiv16}. The theoretical frameworks, based on a variety of approaches (scattering matrices, Green's functions, time-domain calculations, boundary-element method, fluctuating surface and volume currents), share the possibility of addressing bodies of arbitrary geometries and optical properties. Even if some of them can tackle the general scenario of $N$ bodies, so far only a few applications involving more than two bodies have been considered. More specifically, the heat transfer has been analyzed between three nanospheres in the dipolar approximation~\cite{BenAbdallahPRL11,ZhuPRL16}, between three parallel slabs~\cite{ZhengNanoscale11,MessinaPRL12,MessinaPRA14,BenAbdallahPRL14,LatellaPRAppl15}, as well as in a configuration involving one sphere between two slabs~\cite{MullerarXiv}. It has to be mentioned that the radiative heat transfer in networks of more than two particles has recently received a considerable attention, but only within the dipolar approximation~\cite{MessinaPRB13,BenAbdallahPRL13,LanglaisOptExpress14,NikbakhtJAP14,BenAbdallahAPL06,BenAbdallahPRB08,NikbakhtEPL15,BenAbdallahPRL16,LatellaarXiv16}. Concerning the Casimir force, it has been discussed in the case of an atom between two slabs~\cite{MessinaPRA14}, three parallel slabs~\cite{MessinaPRA14}, and very recently the Casimir energy has been considered in the case of two and three coupled cavities when the materials undergo a phase transition from the metallic to the superconducting phase~\cite{Rosa17}.

In this paper, we focus on a system composed of $N$ planar parallel slabs made of arbitrary materials. The $N$ slabs, as well as the external environment in which they are immersed, have arbitrary temperatures. For this scenario, we derive closed-form analytical expressions for the radiative heat transfer and the Casimir force, both at and out of thermal equilibrium. To this aim, we generalize the scattering-matrix approach previously introduced for two~\cite{MessinaEurophysLett11,MessinaPRA11} and three~\cite{MessinaPRA14} bodies. The analytical expressions we obtain clearly highlight the nonadditive character of both momentum and energy exchange. We then consider two numerical applications, one for the Casimir force and one for heat transfer, on a system made of four slabs. For the former, we show how the equilibrium position of the central slab in a system of three slabs is modified by the introduction and lateral shift of a fourth one. For the latter, by fixing some of the temperatures in the system, we discuss the distribution of the other temperatures at local equilibrium as a function of the system parameters.

The paper is structured as follows. In Sec.~\ref{sec:phys_syst}, we present our physical system and introduce the main notation and definitions. Then, in Secs.~\ref{sec:energy_fluxes} and \ref{sec:Casimir}, we formally derive the expressions of the Poynting vector and the stress tensor, respectively. In Sec.~\ref{sec:many_body_sc}, we give explicit expressions of the scattering coefficients, while the energy transmission coefficients and the momentum transfer coefficients are computed in Secs.~\ref{sec:transmission_coefficients} and \ref{sec:casimir_coefficients}, respectively. Then, we present in Sec.~\ref{sec:numerics} our numerical applications to both Casimir force and radiative heat transfer. We finally give some conclusive remarks in Sec.~\ref{sec:conclusions}.

\begin{figure*}
\centering
\includegraphics[scale=1]{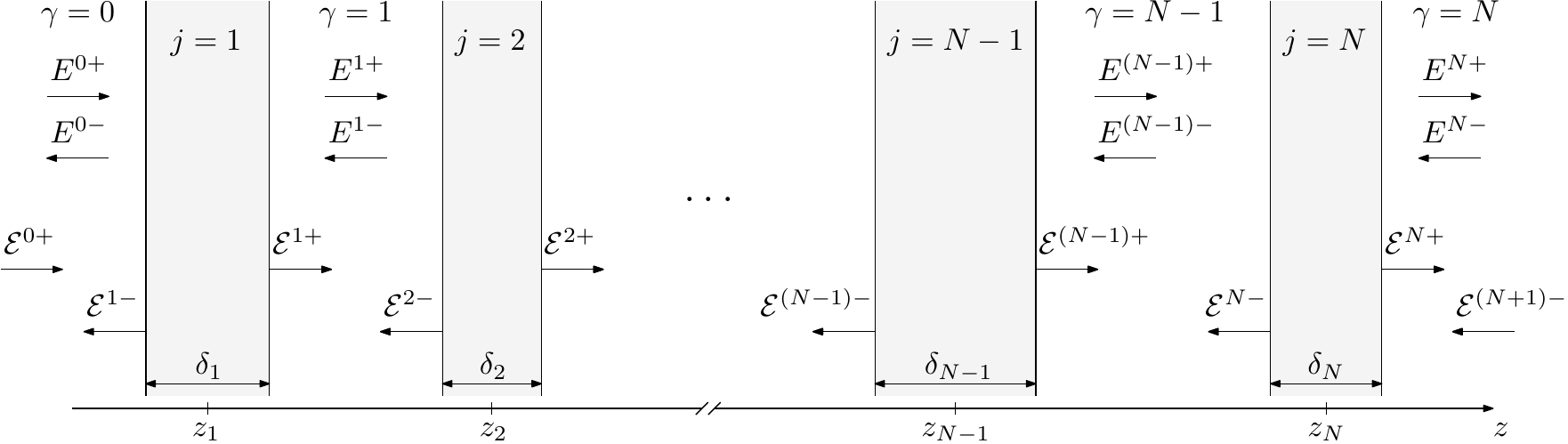} 
\caption{Representation of the $N$-body system. The source fields and the total electric field in each region are also indicated.}
\label{fig1}
\end{figure*}

\section{Many-body systems}
\label{sec:phys_syst}

The system we address is composed of $N$ planar slabs orthogonal to the $z$ axis and assumed to be infinite in the $x$ and $y$ directions. Each slab is at equilibrium at temperature $T_ j$, thermalized by means of some external source, with $ j =1,2,\dots,N$. Let the temperature of the left environment be $T_0$ and that of the right environment be $T_{N+1}$, which can be seen as the equilibrium temperatures of two blackbodies at infinity to which we assign the labels $ j =0$ and $ j =N+1$, respectively. Hereafter we refer to this configuration as an $N$-body system. This distribution of bodies defines $N+1$ vacuum regions that we denote by $\gamma=0,1,\dots,N$. Moreover, for $ j =1,\dots,N$, the $ j $-th slab has thickness $\delta_ j$ and is centered at $z_ j$ on the $z$ axis, as shown in Fig.~\ref{fig1}.

We want to compute energy and momentum fluxes across a surface within any of the vacuum regions of the system. The electric field at a point $\vect{R}=(x,y,z)$ and time $t$ in region $\gamma$, which is created by the fluctuating currents inside the materials, can be expressed as a Fourier expansion given by
\begin{equation}
\vect{E}^{\gamma}(\vect{R},t)=\int_{-\infty}^\infty\frac{\dif\omega}{2\pi}\,e^{-i\omega t}\,\vect{E}^{\gamma}(\vect{R},\omega), 
\end{equation}
where $\omega$ is the frequency. We require that $\vect{E}^{\gamma}(\vect{R},-\omega)= {\vect{E}}^{\gamma *}(\vect{R},\omega)$ in order for the field $\vect{E}^{\gamma}(\vect{R},t)$ to be real, where the asterisk denotes complex conjugation.
In addition, the components $\vect{E}^{\gamma}(\vect{R},\omega)$ can be decomposed using a plane-wave description~\cite{MessinaPRA14}, in such a way that a single mode of the field is specified by the frequency $\omega$, the component $\vect{ k }=(k_x,k_y)$ of the wave vector on the $x$-$y$ plane, the two polarizations $p=\mathrm{TE},\mathrm{TM}$, and the direction of propagation along the $z$ axis which is denoted by $\phi$. The latter can take two values: $\phi=+$ indicating propagation to the right and $\phi=-$ that indicates propagation to the left. The total wave vector reads $\vect{K}^\phi=(\vect{ k },\phi k_z)$, where the component $k_z$ is given by
\begin{equation}
k_z =\sqrt{\frac{\omega^2}{c^2}- k ^2},
\end{equation} 
$c$ being the speed of light in vacuum. We note that when $ k =|\vect{ k }|\leq\omega/c$, the component $k_z$ is real and, therefore, the wave is propagative. When $ k >\omega/c$, $k_z$ is imaginary and the associated wave is evanescent. Evanescent waves are nonpropagating modes of the field: for these modes $\phi$ indicates the direction along which the amplitude of the wave decays. 
Thus, taking this plane-wave decomposition into account, the single-frequency component of the electric field can be written as
\begin{equation}
\vect{E}^{\gamma}(\vect{R},\omega)=\int\frac{\dif^2\vect{  k  }}{(2\pi)^2}\sum_{p,\phi}e^{i\vect{K}^\phi \cdot \vect{R}}\,\hat{\vect{\epsilon}}^\phi(\omega,\vect{ k },p) E^{\gamma\phi}(\omega,\vect{ k },p),
\label{electric_field}
\end{equation}
where $E^{\gamma\phi}(\omega,\vect{ k },p)$ are the components of the electric field in this decomposition and $\hat{\vect{\epsilon}}^\phi(\omega,\vect{ k },p)$ are the polarization vectors. Besides, using the Maxwell equation 
$\nabla\times \vect{E}^{\gamma}(\vect{R},t) =-\partial_t \vect{B}^{\gamma}(\vect{R},t)$, the magnetic field $\vect{B}^{\gamma}(\vect{R},t)$ can be obtained from the electric field and hence, the Fourier components $\vect{B}^{\gamma}(\vect{R},\omega)$ can be expanded in terms of the plane-wave components $E^{\gamma\phi}$. 

Each mode of the total field $E^{\gamma\phi}$ in any region $\gamma$ depends on the fields generated by all the bodies as well as on the background fields present in the left and right environments. More specifically, the mode $E^{\gamma\phi}$ in region $\gamma$ results from the source field modes that we denote (see Fig.~\ref{fig1}) with $\mathcal{E}^{ j \phi}=\mathcal{E}^{ j \phi}(\omega,\vect{ k },p)$, $ j =0,\dots,N+1$. At this stage we observe that the fact of considering only parallel planar slabs introduces a major simplification in the calculation. As a matter of fact, the stationarity of the problem, as well as the translational invariance along the $x$ and $y$ axes safely allow us to state that the frequency $\omega$, the wave vector $\vect{k}$ and the polarization $p$ are conserved in any scattering process. Moreover, thanks to the cylindrical symmetry with respect to the $z$ axis, any reflection and transmission coefficient will depend only on the modulus of the wave vector. As a result, the total field in each region can be written as a linear combination of the form
\begin{equation}
E^{\gamma\phi}=\sum_{ j ,\eta}L^{\gamma\phi}_{ j \eta} \mathcal{E}^{ j \eta},
\label{coefficients_L_0}
\end{equation}
in terms of the coefficients $L^{\gamma\phi}_{ j \eta}=L^{\gamma\phi}_{ j \eta}(\omega, k ,p)$. 
Hereafter, if not otherwise explicitly stated, summations over indices labeling bodies run from $0$ to $N+1$, including quantities associated with the environmental fields. Equation \eqref{coefficients_L_0} will allow us in the following to relate the statistical properties of the total field in each region, needed to calculate both the Casimir force and the heat transfer, to the statistical properties of the individual source fields, directly derived from the fluctuation-dissipation theorem.

\section{Radiative heat transfer}
\label{sec:energy_fluxes}

We can now move toward an explicit expression of the radiative heat transfer on each slab. To this aim, we should first observe that our assumption of infinite extension of all the slabs actually leads to a formally infinite flux. Of course, the same issue applies to Casimir force as well. Nevertheless, the translational invariance allows us to think in terms of flux and force per unit surface. As discussed in detail in Ref.~\cite{MessinaPRA11}, the transition from the calculation of the total flux (or force) to the one per unit surface basically consists of omitting an infinite Dirac delta $\delta(\mathbf{0})$ and its coefficient $(2\pi)^2$ in the final expressions.

Starting from the case of heat transfer, the energy flux per unit surface in region $\gamma$ is given by the averaged $z$ component of the Poynting vector $\vect{S}^{\gamma}(\vect{R},t)$, calculated at $z=\bar{z}_\gamma$, where $\bar{z}_\gamma$ is located in zone $\gamma$. In Cartesian components, we have
\begin{equation}
\left\langle S^{\gamma}_i(\vect{R},t)\right\rangle=\varepsilon_0c^2\sum_{j,k}\epsilon_{ijk}\left\langle E^{\gamma}_j(\vect{R},t)B^{\gamma}_k(\vect{R},t)\right\rangle, 
\label{Poynting_vector}
\end{equation} 
where $\left\langle\; \cdots\right\rangle$ indicates symmetrized statistical average, $\epsilon_{ijk}$ is the Levi-Civita symbol with $i,j,k=x,y,z$, and $\varepsilon_0$ is the vacuum permittivity. In order to compute this quantity, we introduce the correlation functions $C^{\gamma\phi\phi'}=C^{\gamma\phi\phi'}(\omega, k ,p)$ which are defined according to
\begin{equation}\begin{split}
\big\langle  E^{\gamma\phi}(\omega,\vect{ k },p)E^{\gamma\phi'\dag}(\omega',\vect{ k }',p') \big\rangle&=  (2\pi)^3 \delta\left(\omega-\omega'\right)\\
&\,\times\delta\left(\vect{ k }-\vect{ k }'\right) \delta_{pp'} C^{\gamma\phi\phi'},
\end{split}\label{correlation_functions}\end{equation}
where the dagger denotes hermitian conjugate. Thus, using Eqs.~\eqref{Poynting_vector} and \eqref{correlation_functions}, after manipulating the polarization vectors, the averaged $z$ component of the Poynting vector takes the form~\cite{MessinaPRA14}
\begin{equation}\begin{split}
\Phi_{\gamma}&\equiv\left\langle S^{\gamma}_z(\vect{R},t)\right\rangle\\
&=\int_0^\infty\frac{\dif\omega}{2\pi}\int_0^\infty\frac{\dif  k }{2\pi}\,  k \sum_{p,\phi,\phi'}
2\varepsilon_0c^2\frac{k_z }{\omega} \phi\\
&\,\times\left[\Pi^\mathrm{pw}\delta_{\phi\phi'}+\Pi^\mathrm{ew}\left(1-\delta_{\phi\phi'}\right)\right]C^{\gamma\phi\phi'},
\end{split}\label{energy_flux}\end{equation}
where we have introduced the projectors on the propagating and evanescent wave sectors $\Pi^\mathrm{pw}$ and $\Pi^\mathrm{ew}$, respectively, defined by
\begin{equation}
\Pi^\mathrm{pw}\equiv\theta(\omega-c k ),\qquad
\Pi^\mathrm{ew}\equiv\theta(c k -\omega),
\end{equation}
$\theta(x)$ being the Heaviside step function.
We observe that, as a consequence of Eq.~\eqref{correlation_functions}, the energy flux $\Phi_{\gamma}$ is stationary and invariant under translations in the $x$-$y$ plane. Notice that the dependence on $z$ in Eq.~\eqref{energy_flux} is implicit through the correlation functions of the total field in a particular region $\gamma$ of the system. Furthermore, we already have an expression [Eq.~\eqref{coefficients_L_0}] connecting the total field in each region to the fields emitted by the bodies and the environments. We now perform the so-called local-thermal-equilibrium approximation, i.e. we assume that each body radiates as it would do at equilibrium at its own temperature, so that the modes of the source fields corresponding to different bodies are not correlated to each other. Denoting by $\mathcal{C}^{ j \eta\eta'}=\mathcal{C}^{ j \eta\eta'}(\omega, k ,p)$ the correlation functions of the source fields associated to body $ j $, with $\eta,\eta'=+,-$, the assumption of local thermal equilibrium leads to~\cite{MessinaPRA14}
\begin{equation}
C^{\gamma\phi\phi'}=\sum_{ j ,\eta,\eta'} L^{\gamma\phi}_{ j \eta} L^{\gamma\phi'*}_{ j \eta'} \mathcal{C}^{ j \eta\eta'}.
\label{correlation_total_field}
\end{equation}
In addition, for convenience, we introduce the coefficients $\mathcal{K}_{\phi\phi'}^{ j \eta\eta'}=\mathcal{K}_{\phi\phi'}^{ j \eta\eta'}(\omega, k ,p)$ defined according to
\begin{equation}
\left[\Pi^\mathrm{pw}\delta_{\phi\phi'}+\Pi^\mathrm{ew}\left(1-\delta_{\phi\phi'}\right)\right]\mathcal{C}^{ j \eta\eta'}
\equiv\frac{\hbar\omega^2 \mathcal{N}_ j }{2\varepsilon_0c^2k_z} \mathcal{K}_{\phi\phi'}^{ j \eta\eta'},
\label{coefficients_K}
\end{equation}
where
\begin{equation}\begin{split}
\mathcal{N}_ j (\omega)&= n_ j (\omega)+\frac{1}{2}, \label{Theta}\\
n_ j (\omega)&=\left(e^{\hbar\omega/\kB T_ j  }-1\right)^{-1},
\end{split}\end{equation}
$\kB$ is the Boltzmann constant, and $\hbar$ the reduced Planck constant.
Taking into account the previous definitions, the energy flux \eqref{energy_flux} becomes
\begin{equation}
\Phi_{\gamma}=\int_0^\infty\frac{\dif\omega}{2\pi}\int_0^\infty\frac{\dif  k }{2\pi}\,  k \sum_{p, j }\hbar\omega\mathcal{N}_ j  X^{\gamma, j },
\label{energy_flux_2}
\end{equation}
where the coefficients $X^{\gamma, j }=X^{\gamma, j }(\omega, k ,p)$ are given by
\begin{equation}
X^{\gamma, j } = \sum_{\phi,\phi',\eta,\eta'}\phi L^{\gamma\phi}_{ j \eta} L^{\gamma\phi'*}_{ j \eta'}  \mathcal{K}^{ j \eta\eta'}_{\phi\phi'}.
\label{coefficients_X}
\end{equation}

We note that the dependence on the temperature in Eq.~\eqref{energy_flux_2} is explicit through the functions $\mathcal{N}_ j $ and possibly implicit in the optical properties (and thus in the scattering amplitudes) of the bodies. We show in Appendix~\ref{derivation_eq_X} that the above coefficients always satisfy
\begin{equation}
\sum_ j  X^{\gamma, j }=0.
\label{equation_X}
\end{equation}
As a consequence, if all the bodies are thermalized at the same equilibrium temperature, using Eq.~\eqref{equation_X} in Eq.~\eqref{energy_flux_2} one sees that the flux $\Phi_{\gamma}$ vanishes in each region $\gamma$.

In view of Eq.~\eqref{equation_X}, we observe here that $\sum_ j \mathcal{N}_ j  X^{\gamma, j }=\sum_ j  n_ j  X^{\gamma, j }$ and, therefore, purely quantum contributions associated to zero-point fluctuations do not participate in the energy fluxes.
Furthermore, since the net energy flux on body $ j =1,\dots,N$ is given by $\Phi^{ j }=\Phi_{ j -1}-\Phi_{ j }$, taking into account Eq.~\eqref{equation_X} this flux can be written as
\begin{equation}
\Phi^{ j }=\int_0^\infty\frac{\dif\omega}{2\pi}\int_0^\infty\frac{\dif  k }{2\pi}\,  k \sum_{p}\sum_{ \ell \neq j }
\hbar\omega n_{ \ell , j } \mathcal{T}^{ \ell , j },
\label{total_energy_flux_2}
\end{equation}
where we have introduced $n_{ \ell , j }\equiv n_ \ell -n_ j $ and the energy transmission coefficients $\mathcal{T}^{ \ell , j }=\mathcal{T}^{ \ell , j }(\omega, k ,p)$ given by
\begin{equation}
\mathcal{T}^{ \ell , j }= X^{ j -1, \ell }-X^{ j , \ell }, \qquad j=1,\dots,N.
\label{trans_coeff}
\end{equation}
In this way, as can be seen from Eq.~\eqref{total_energy_flux_2}, energy fluxes are described with a Landauer-like formalism in many-body systems. The above energy transmission coefficients will be computed in Sec.~\ref{sec:transmission_coefficients}, where, in particular, it can be seen that they satisfy the reciprocity relation $\mathcal{T}^{ \ell , j }=\mathcal{T}^{ j , \ell }$, with $ j , \ell =1,\dots,N$.

We highlight that as a consequence of Eq.~\eqref{equation_X} and the definition \eqref{trans_coeff}, the energy transmission coefficients satisfy the following remarkable property: $\sum_\ell \mathcal{T}^{\ell, j}=0$. On the one hand, using this result one immediately obtains $\mathcal{T}^{j, j}= -\sum_{\ell \neq j} \mathcal{T}^{\ell, j}$. Since $\mathcal{T}^{\ell,j}$ with $\ell \neq j$ are positive quantities, the coefficients $\mathcal{T}^{j, j}$ are negative. On the other hand, consider now a situation in which all the bodies are assumed to be thermalized at $T_\ell = 0$, except body $j$ for which $T_j > 0$. Under these conditions and according to the previous reasoning, from Eq.~\eqref{total_energy_flux_2}, the net energy flux on body $j$ can be expressed as
\begin{equation}
\Phi^{ j }= \int_0^\infty\frac{\dif\omega}{2\pi}\int_0^\infty\frac{\dif  k }{2\pi}\,  k \sum_{p} \hbar\omega n_{ j } \mathcal{T}^{ j , j }.
\label{total_energy_flux_3}
\end{equation}
Notice that $\Phi^j < 0$. The above expression allows us to interpret the coefficient $\mathcal{T}^{j , j}$ as the self-emission amplitude associated to body $j$, since it accounts for the radiation emitted by this body in the presence of the rest of the system. In this case, $\Phi^j$ corresponds to the self-emission rate discussed in Ref.~\cite{MullerarXiv}.

\section{Casimir-Lifshitz forces}
\label{sec:Casimir}

We now focus on the Casimir-Lifshitz forces acting on the system. The formulation we introduce here is analogous to the previous one for heat transfer, but now momentum fluxes are the relevant quantities to be considered instead of energy fluxes. Since energy and momentum fluxes are quantities with different physical nature, however, a priori one expects some differences to arise when comparing the two descriptions. For instance, the momentum transfer coefficients we introduce below are formally analogous to the energy transmission coefficients, but the former are represented by complex numbers while the latter are real. This fact is easy to understand if one bears in mind that the energy quanta $\hbar\omega$ are always real, whereas the momentum component $\hbar k_z$ becomes imaginary for photons characterizing evanescent fields. Another important difference is that nonvanishing momentum fluxes occur even at thermal equilibrium and, as is well known, purely quantum fluctuations contribute to these fluxes as well. As a consequence, reciprocity will not hold for these momentum transfer coefficients.

To describe Casimir-Lifshitz forces, let us consider the Maxwell stress tensor $\mathbb{T}^{\gamma}(\vect{R},t)$ in a particular region $\gamma$ of the system, whose Cartesian components read
\begin{equation}\begin{split}
T^{\gamma}_{ij}(\vect{R},t)&= \varepsilon_0\left[E_i^{\gamma}(\vect{R},t)E_j^{\gamma}(\vect{R},t)
+c^2B_i^{\gamma}(\vect{R},t)B_j^{\gamma}(\vect{R},t)\right] \\
&\,-\frac{\varepsilon_0}{2}\delta_{ij}\left[\left|\vect{E}^{\gamma}(\vect{R},t)\right|^2
+c^2\left|\vect{B}^{\gamma}(\vect{R},t)\right|^2\right]
\end{split}\end{equation}
with $i,j=x,y,z$. The momentum flux in region $\gamma$ is given by the averaged component $T^{\gamma}_{zz}(\vect{R},t)$ and takes the form~\cite{MessinaPRA14}
\begin{equation}\begin{split}
P_\gamma&\equiv\left\langle T^{\gamma}_{zz}(\vect{R},t)\right\rangle\\
&=- \int_0^\infty\frac{\dif\omega}{2\pi} 
\int_0^\infty\frac{\dif  k }{2\pi}\,  k  \sum_{p,\phi,\phi'}
2\varepsilon_0 \frac{c^2k_z^2}{\omega^2}\\
&\,\times\left[\Pi^\mathrm{pw}\delta_{\phi\phi'}+\Pi^\mathrm{ew}\left(1-\delta_{\phi\phi'}\right)\right]
C^{\gamma\phi\phi'}.
\end{split}\label{P_gamma_0}\end{equation}
As for the case of heat transfer, the above expression is obtained by expanding the electric field in the plane-wave representation \eqref{electric_field}, using Maxwell's equations to obtain the magnetic field in terms of the electric plane-wave components, manipulating the polarization vectors, and introducing the correlation functions with the help of Eq.~\eqref{correlation_functions}. Moreover, also here we will take into account the local-equilibrium approximation to write the correlation functions of the total field in terms of the correlation functions of the source fields. According to this, using Eqs.~\eqref{correlation_total_field} and \eqref{coefficients_K}, Eq.~\eqref{P_gamma_0} can be rewritten as
\begin{equation}
P_\gamma = - \int_0^\infty\frac{\dif\omega}{2\pi}
\int_0^\infty\frac{\dif  k }{2\pi}\,  k  \sum_{p, j } \hbar k_z \mathcal{N}_ j 
Y^{\gamma, j },
\label{P_gamma_1}
\end{equation} 
where the coefficients $Y^{\gamma, j }=Y^{\gamma, j }(\omega, k ,p)$ are given by
\begin{equation}
Y^{\gamma, j }= \sum_{\phi,\phi',\eta,\eta'} L^{\gamma\phi}_{ j \eta} L^{\gamma\phi'*}_{ j \eta'}  \mathcal{K}^{ j \eta\eta'}_{\phi\phi'}.
\label{coefficients_Y}
\end{equation}

The net force per unit area acting on body $ j =1,\dots,N$ can be computed as $P^ j =P_ j -P_{ j -1}$, so that using Eq.~\eqref{P_gamma_1} we can write
\begin{equation}
P^ j = \int_0^\infty\frac{\dif\omega}{2\pi}
\int_0^\infty\frac{\dif  k }{2\pi}\,  k  \sum_{p, \ell } \hbar k_z \mathcal{N}_ \ell  \mathcal{W}^{ \ell , j },
\label{net_force}
\end{equation}
where we have introduced the previously mentioned momentum transfer coefficients $\mathcal{W}^{ \ell , j }=\mathcal{W}^{ \ell , j }(\omega, k ,p)$ defined by
\begin{equation}
\mathcal{W}^{ \ell , j }= Y^{ j -1, \ell }-Y^{ j , \ell },\qquad j =1,\dots,N.
\label{W}
\end{equation}
In addition, the previous expression of the net pressure $P^j$, Eq.~\eqref{net_force}, can be conveniently rewritten as
\begin{equation}
\begin{split}
P^ j &= \int_0^\infty\frac{\dif\omega}{2\pi}\int_0^\infty\frac{\dif  k }{2\pi}\,  k  \sum_p \hbar k_z \mathcal{N}_ j  \sum_{ \ell } \mathcal{W}^{ \ell , j }\\
&+\int_0^\infty\frac{\dif\omega}{2\pi}\int_0^\infty\frac{\dif  k }{2\pi}\,  k \sum_{p} \sum_{ \ell \neq j } \hbar k_z n_{ \ell , j }\mathcal{W}^{ \ell , j }.
\end{split} 
\label{net_force_2}
\end{equation}
Hence, in particular, we see that at thermal equilibrium at temperature $T=T_j$, the functions $n_{\ell,j}$ in the second term of Eq.~\eqref{net_force_2} vanish and, therefore, the net pressure on body $j$ reduces to
\begin{equation}
P^ j _\mathrm{eq}= \int_0^\infty\frac{\dif\omega}{2\pi}
\int_0^\infty\frac{\dif  k }{2\pi}\,  k  \sum_p \hbar k_z \mathcal{N}_ j  \sum_{ \ell } \mathcal{W}^{ \ell , j }.
\label{net_force_equilibrium}
\end{equation}

The momentum transfer coefficients $\mathcal{W}^{ \ell , j }$ will be given in Sec.~\ref{sec:casimir_coefficients}. To obtain these coefficients (and the energy transmission coefficients), in the next section we first introduce the tools needed to solve the many-body scattering problem in terms of single-body reflection and transmission coefficients.

\section{Scattering and electric field coefficients}\label{sec:many_body_sc}

Having introduced a formal method to compute the couplings driving energy and momentum exchanges, a procedure to relate such couplings to individual properties is required. To this aim, below we start by considering a systematic way to characterize many-body scattering processes in terms of optical properties of the single constituents of the system. Using this procedure, we subsequently determine the electric field coefficients $L^{\gamma\phi}_{ j \eta}$ which are necessary to express energy transmission and momentum transfer coefficients.

\subsection{Many-body scattering coefficients}

The scattering operators introduced in Refs.~\cite{MessinaPRA11,MessinaPRA14} for two- and three-body systems, which for our geometry reduce simply to coefficients, are a useful tool that permits us to write physical quantities in a convenient way. Our aim here is to introduce such coefficients for the $N$-body case we are concerned with.

The many-body scattering coefficients take into account the presence of different bodies at the same time and are built in terms of the single-body reflection and transmission coefficients denoted by
$\rho^{ j }_\phi=\rho^{ j }_\phi(\omega, k ,p)$ and $\tau^{ j }=\tau^{ j }(\omega, k ,p)$, respectively, where $ j $ labels the associated body. For the reflection coefficients, $\phi=+,-$ specify the direction of propagation or decay of the outgoing field (the incoming field propagates or decays in the direction $-\phi$), whereas the transmission coefficients do not depend on $\phi$. 
On the one hand, for $ j =1,\dots,N$, these coefficients can be written as
\begin{equation}\begin{split}
\rho^{ j }_\phi&=\rho_{ j } e^{-i k_z(\delta_ j +\phi2z_ j )},\\
\tau^{ j } &= \tau_{ j } e^{-i k_z \delta_ j },
\end{split}\end{equation}
with $\rho_{ j }=\rho_{ j }(\omega, k ,p,\delta_ j )$ and $\tau_{ j }=\tau_{ j }(\omega, k ,p,\delta_ j )$ depending implicitly on the optical properties of the single body.
For the geometry under consideration and for isotropic media, the latter take the form
\begin{equation}\begin{split}
\rho_{ j }&=r_{p, j }\frac{1-e^{2i k_{z  j }\delta_ j }}{1-r^2_{p, j }
e^{2i k_{z  j }\delta_ j }},\\
\tau_{ j }&=\frac{\left(1- r_{p, j }^2\right)
e^{i k_{z  j }\delta_ j }}
{1-r^2_{p, j }e^{2i k_{z  j }\delta_ j }}.
\end{split}\end{equation}
Here the $z$ component of the wave vector inside medium $ j $ reads
\begin{equation}
k_{z  j }=\sqrt{\frac{\omega^2}{c^2}\varepsilon_ j \mu_ j - k ^2},
\end{equation}
and the vacuum-medium Fresnel reflection coefficients (for $p=\mathrm{TE},\mathrm{TM}$) are given by
\begin{equation}
r_{\mathrm{TE},  j }=\frac{\mu_ j  k_z-k_{z  j }}{\mu_ j  k_z+k_{z  j }}, \qquad
r_{\mathrm{TM},  j }=\frac{\varepsilon_ j  k_z-k_{z  j }}{\varepsilon_ j  k_z+k_{z  j }},
\label{Fresnel_coefficients}
\end{equation}
where $\varepsilon_ j $ and $\mu_ j $ are the dielectric permittivity and magnetic permeability of body $ j $, respectively. 
On the other hand, for the blackbodies radiating the environmental fields ($ j =0$ and $ j =N+1$) we set 
\begin{equation}\begin{split}
\rho^{0}_\phi&=\rho^{N+1}_\phi=0,\\
\tau^{0}&=\tau^{N+1}=0.
\end{split}\label{reflection_transmission_boundary}\end{equation}

In addition, the correlation functions of the source fields, including those of the environments, are defined in terms of the previously introduced single-body reflection and transmission coefficients. These correlation functions are given by~\cite{MessinaPRA11}
\begin{equation}\begin{split}
\mathcal{C}^{ j \eta\eta'}&= \frac{\hbar\omega^2\mathcal{N}_ j }{2\varepsilon_0c^2k_z} \big\{\delta_{\eta\eta'}\\
&\,\times \big[ \Pi^{\text{pw}} \big(1-\big|\rho_\eta^ j \big|^2 
-\big|\tau^{ j }\big|^2\big)
+ \Pi^{\text{ew}} \big(\rho_\eta^ j  -\rho_\eta^{ j *}\big) \big]\\
&\,+ \big(1-\delta_{\eta\eta'}\big) \big[\Pi^{\text{pw}} \big(-\rho_\eta^ j \tau^{ j *}-\rho_{\eta'}^{ j *}\tau^{ j }\big)\\
&\,+ \Pi^{\text{ew}}\big(\tau^{ j }-\tau^{ j *}\big) \big] \big\}.
\end{split}\label{correlations_source_fields}\end{equation}
Using this result, the coefficients $\mathcal{K}^{ j \eta\eta'}_{\phi\phi'}$ introduced in Eq.~\eqref{coefficients_K} can be easily computed and expressed as
\begin{equation}\begin{split}
\mathcal{K}^{ j \eta\eta'}_{\phi\phi'}
&=\Pi^{\text{pw}}\delta_{\phi\phi'}
\big[\delta_{\eta\eta'}\big(1-\big|\rho_\eta^ j \big|^2-\big|\tau^{ j }\big|^2\big)\\
&\,- \big(1-\delta_{\eta\eta'}\big) \big(\rho_\eta^ j  \tau^{ j *}
+\rho_{\eta'}^{ j *}\tau^{ j }\big) \big]\\
&\,+ \Pi^{\text{ew}}\big(1-\delta_{\phi\phi'}\big) 
\big[\delta_{\eta\eta'}\left(\rho_\eta^ j -\rho_\eta^{ j *}\right)\\
&\,+ \big(1-\delta_{\eta\eta'}\big)\big(\tau^{ j }-\tau^{ j *}\big)\big].
\end{split}\label{coefficients_K_2}\end{equation} 

In order to define the scattering coefficients for the $N$-body case, consider a block of consecutive bodies having indexes going from $j$ to $m$ (with $ j\leq m$), and let us denote the sequence of these bodies by $ j \to m$. The reflection and transmission coefficients for this block, $\rho_\phi^{ j \to m}$ and $\tau^{ j \to m}$, representing the analogues of $\rho_\phi^{ j }$ and $\tau^{ j }$ for a single body, are given by
\begin{equation}\begin{split}
\rho_+^{ j \to m}&=\rho_+^{ \ell\to m} +\left(\tau^{ \ell\to m}\right)^2 u^{ j \to \ell-1, \ell\to m}\rho_+^{ j \to \ell-1},\\
\rho_-^{ j \to m}&=\rho_-^{ j \to \ell-1} + \left(\tau^{ j \to \ell-1}\right)^2 u^{ j \to \ell-1, \ell\to m}\rho_-^{ \ell\to m},\\
\tau^{ j \to m}&=\tau^{ j \to \ell-1}u^{ j \to \ell-1, \ell\to m} \tau^{ \ell\to m} ,
\end{split}\label{many-body_sc}\end{equation}
where $j< \ell\leq m$ and
\begin{equation}\begin{split}
u^{ j \to \ell-1, \ell\to m}
&=\sum_{n=0}^\infty\left(\rho_+^{ j\to \ell-1}\rho_-^{ \ell\to m}\right)^n \\
&=\left(1-\rho_+^{ j \to \ell-1}\rho_-^{ \ell\to m}\right)^{-1}.
\end{split}\label{Fabry-Perot}\end{equation}
The coefficient $\rho_+^{ j \to m}$ as expressed in Eq.~\eqref{many-body_sc}, for example, accounts for the reflection of a mode to the right due to bodies $ j \to m$ together; it has a direct contribution from the reflection produced by bodies $ \ell\to m$, and a contribution that takes into account that the mode is transmitted through bodies $ \ell\to m$, undergoes multiple reflections within the cavity formed by bodies $ j \to \ell-1$ and $ \ell\to m$, is reflected to the right by bodies $ j \to \ell-1$, and finally leaves the cavity by transmission through bodies $ \ell\to m$. Analogously, the coefficient $\tau^{ j \to m}$ given in Eq.~\eqref{many-body_sc} represents transmission through bodies $ j \to \ell-1$, multiple reflections between bodies $ j \to \ell-1$ and $ \ell\to m$, and transmission through bodies $ \ell\to m$.

According to the above expressions, the many-body scattering coefficients can be equivalently computed in several ways by choosing different allowed values of $\ell$. For convenience, however, below we introduce a setup that is particularly useful for the problem at hand. We rewrite these coefficients as
\begin{equation}\begin{split}
\rho_+^{ j \to m}&= \hat{\rho}_+^{ j \to m}e^{-i k_z\left(\delta_m+2z_m\right)},\\
\rho_-^{ j \to m}&= \hat{\rho}_-^{ j \to m}e^{-i k_z\left(\delta_j-2z_j\right)},\\
\tau^{ j \to m}&= \hat{\tau}^{ j \to m}\exp\Bigl(-i k_z \sum_{\ell=j}^m \delta_\ell\Bigr),
\end{split}\label{hat_coefficients}\end{equation}
where from Eq.~\eqref{many-body_sc} we have, for $m>j$,
\begin{equation}\begin{split}
\hat{\rho}_+^{ j \to m}&= \rho_m + (\tau_m)^2 \hat{\rho}_+^{ j \to m-1}u^{ j \to m-1, m} e^{2i k_z d_{ m-1}} ,\\
\hat{\rho}_-^{ j \to m}&= \rho_j + (\tau_j)^2 \hat{\rho}_-^{ j +1\to m}  u^{ j , j +1\to m} e^{2i k_zd_j }, \\
\hat{\tau}^{ j \to m}&= \hat{\tau}^{ j \to m-1} u^{ j \to m-1, m}  \tau_m,
\end{split}\end{equation}
and
\begin{equation}\begin{split}
u^{ j \to m-1, m}&= \left(1-\hat{\rho}_+^{ j \to m-1} \rho_m e^{2i k_zd_{ m-1}} \right)^{-1},\\
u^{ j , j +1\to m}&= \left(1-\rho_j \hat{\rho}_-^{ j +1\to m} e^{2i k_zd_j}\right)^{-1}.
\end{split}\end{equation}
Here $d_{ j }$ is the separation distance between the consecutive bodies $ j $ and $ j  +1$, given by
\begin{equation}
d_{ j }= z_{ j +1}-z_{ j }-\delta_{ j }/2-\delta_{ j +1}/2,
\label{separation}
\end{equation}
and which corresponds to the width of region $\gamma= j $. Notice that the above coefficients are to be taken as $\hat{\rho}_\phi^{ j }=\rho_ j $ and $\hat{\tau}^{ j }=\tau_ j $ for a single body.

\subsection{Electric field coefficients}

To obtain the expressions for the energy transmission and momentum transfer coefficients, we have to determine first the coefficients $L_{ j \eta}^{\gamma\phi}$ relating source fields to total fields in a given region of the system. As stated by Eq.~\eqref{coefficients_L_0}, the coefficients $L_{ j \eta}^{\gamma\phi}$ account for the contribution of the field mode $\mathcal{E}^{ j \eta}$, emitted by the source $j$ in direction $\eta$, to the total field mode $E^{\gamma\phi}$ in region $\gamma$ and direction $\phi$. On this basis and using the many-body scattering coefficients, we are able to directly write down some useful relations between these coefficients that will allow us to find $L_{ j \eta}^{\gamma\phi}$. For instance, recalling that $\gamma=0,\dots,N$ and $ j =0,\dots,N+1$, we can write
\begin{equation}\begin{split}
L_{ j \eta}^{\gamma-}&=\rho_-^{\gamma+1\to N+1} L_{ j \eta}^{\gamma+},\qquad  j \leq\gamma,\\ 
L_{ j \eta}^{\gamma+}&=\rho_+^{0\to\gamma} L_{ j \eta}^{\gamma-},\qquad  j >\gamma.
\end{split}\label{L13}\end{equation}
The first of these relations indicates that the contribution of the source mode $\mathcal{E}^{ j \eta}$ to the total mode $E^{\gamma-}$ is proportional to the contribution of the same source mode to the total mode $E^{\gamma+}$. If the source is located on the left of the considered region ($j \leq\gamma$), the proportionality factor is the backward reflection coefficient $\rho_-^{\gamma+1\to N+1}$ of the block formed by all the bodies on the right of the region (see Fig.~\ref{fig1} for illustration).
An analogous reasoning applies to the second of Eqs.~\eqref{L13}.

We now want to relate $L_{ j -}^{\gamma+}$ to $L_{ j +}^{\gamma+}$ for $j \leq\gamma$. On the one hand, since the (left) environment $j=0$ only radiates to the right, the coefficient $L_{ j -}^{\gamma+}$ must vanish for $j=0$. In addition, this coefficient also vanishes for $j=1$, because the field emitted by this source to the left is not reflected back into the system but is absorbed by the environment. On the other hand, for sources such that $1< j \leq\gamma$, one realizes that a mode emitted to the right by the source $j$ is equivalent to a mode emitted by this source to the left undergoing the following scattering process: multiple reflections in the cavity formed by bodies $0 \to j-1$ and body $j$, reflection to the right by bodies $0 \to j-1$, and transmission through body $j$. Taking this into account, we thus can write   
\begin{equation}
L_{ j -}^{\gamma+}= 
\begin{cases}
0, &  j =0\\
\rho_+^{0\to j -1}
u^{0\to j -1, j }
\tau^{ j } L_{ j +}^{\gamma+},& 0< j \leq\gamma. 
\end{cases}
\label{L5}
\end{equation}
Notice that $\rho_+^{0\to j -1}$ vanishes when $j=1$, so in this case, as previously explained, $L_{ j -}^{\gamma+}$ vanishes as well. 
Following similar arguments one can also deduce that
\begin{equation}
L_{ j +}^{\gamma-}= 
\begin{cases}
\tau^{ j }
u^{ j , j +1\to N+1} 
\rho_-^{ j +1\to N+1} L_{ j -}^{\gamma-}, & \gamma< j  \leq N \\
0, &  j =N+1 
\end{cases}
\label{L6}.
\end{equation}

We highlight at this point that in virtue of properties \eqref{L13}, \eqref{L5}, and \eqref{L6}, the coefficients $L_{ j \eta}^{\gamma\phi}$ are completely determined if $L_{ j +}^{\gamma+}$ and $L_{ j -}^{\gamma-}$ are explicitly known for $j \leq \gamma$ and for $j> \gamma$, respectively. Again, these coefficients can be easily obtained using the many-body scattering coefficients; let us focus on $L_{ j +}^{\gamma+}$ for $j \leq \gamma$. We first note that when $j=\gamma$, the contribution of mode $\mathcal{E}^{ j +}$ to the total field $E^{ \gamma +}$ is simply given by the factor accounting for multiple reflections in the cavity formed by bodies $0\to\gamma$ and bodies $\gamma+1\to N+1$, i.e. $u^{0\to\gamma,\gamma+1\to N+1}$. When $j < \gamma$, in addition to the previous factor, the contribution of mode $\mathcal{E}^{ j +}$ is affected by a factor $u^{0\to j , j +1\to \gamma} \tau^{ j +1\to\gamma}$ that describes the scattering from region $j$ to the considered region $\gamma$. With this, we obtain
\begin{equation}
L_{ j +}^{\gamma+}=
u^{0\to\gamma,\gamma+1\to N+1}
\begin{cases}
u^{0\to j , j +1\to \gamma} 
\tau^{ j +1\to\gamma} , &  j <\gamma\\
1, &  j =\gamma
\end{cases}.
\label{L_plus_plus}
\end{equation}
Finally, by symmetry or by employing analogous arguments, one arrives at the conclusion that 
\begin{equation}
L_{ j -}^{\gamma-}= 
u^{0\to\gamma,\gamma+1\to N+1}
\begin{cases}
1, & j =\gamma+1\\
u^{\gamma+1\to j -1, j \to N+1} \\
\times \tau^{\gamma+1\to j -1}, & j >\gamma+1
\end{cases} .
\label{L_minus_minus}
\end{equation}

We have determined the coefficients $L_{ j \eta}^{\gamma\phi}$, so we are now able to perform the calculation of energy and momentum fluxes.

\section{Energy transmission coefficients}\label{sec:transmission_coefficients}

In this section we determine the energy transmission coefficients $\mathcal{T}^{ \ell , j }$. To proceed, we introduce the set of coefficients $\hat{\mathcal{T}}^{ j }_{\gamma}=\hat{\mathcal{T}}^{ j }_{\gamma}(\omega, k ,p)$ defined by
\begin{equation}
\hat{\mathcal{T}}^{ j }_{\gamma}\equiv \sum_{ \ell =0}^ j  X^{\gamma, \ell },
\label{transmission_coefficients_1}
\end{equation}
As discussed below, the energy transmission coefficients $\mathcal{T}^{ \ell , j }$ can be fully determined in terms of $\hat{\mathcal{T}}^ j _\gamma$.
In Appendix~\ref{derivation_eq_X} we show that the coefficients $\hat{\mathcal{T}}^ j _\gamma$ take the form ($ j ,\gamma=0,\dots,N$)
\begin{widetext}
\begin{equation}\begin{split}
\hat{\mathcal{T}}^{ j }_{\gamma}
&= 
\frac{\Pi^{\text{pw}} \big|\tau^{ j +1\to\gamma}\big|^2 \big(1 -\big|\rho_+^{0\to j }\big|^2\big)
\big( 1 -\big|\rho_-^{\gamma+1\to N+1}\big|^2 \big)}
{\big|1- \rho_+^{0\to\gamma} \rho_-^{\gamma+1\to N+1} \big|^2 
\big|1- \rho_+^{0\to j } \rho_-^{ j +1\to \gamma}\big|^2} 
+ 
\frac{\Pi^{\text{ew}} 4\big|\tau^{ j +1\to\gamma}\big|^2\text{Im}\big(\rho_+^{0\to j }\big)
\text{Im}\big(\rho_-^{\gamma+1\to N+1}\big)}
{\big|1- \rho_+^{0\to\gamma} \rho_-^{\gamma+1\to N+1} \big|^2 
\big|1- \rho_+^{0\to j } \rho_-^{ j +1\to \gamma}\big|^2},
\qquad  j <\gamma,\\
\hat{\mathcal{T}}^{\gamma}_{\gamma}
&=    
\frac{\Pi^{\text{pw}} \big(1 -\big|\rho_+^{0\to\gamma}\big|^2\big)
\big( 1 -\big|\rho_-^{\gamma+1\to N+1}\big|^2 \big)}
{\big|1- \rho_+^{0\to\gamma} \rho_-^{\gamma+1\to N+1}\big|^2}
+ 
\frac{\Pi^{\text{ew}} 4\text{Im}\big(\rho_+^{0\to\gamma}\big)
\text{Im}\big(\rho_-^{\gamma+1\to N+1}\big)}
{\big|1- \rho_+^{0\to\gamma} \rho_-^{\gamma+1\to N+1}\big|^2},\\
\hat{\mathcal{T}}^{ j }_{\gamma}
&=
\frac{\Pi^{\text{pw}} \big|\tau^{\gamma+1\to j } \big|^2 
\big(1-\big|\rho_+^{0\to\gamma}\big|^2\big)
\big(1-\big| \rho_-^{ j +1\to N+1}\big|^2 \big)}
{\big|1- \rho_+^{0\to j } \rho_-^{ j +1\to N+1}\big|^2
\big|1- \rho_+^{0\to\gamma} \rho_-^{\gamma+1\to  j }\big|^2}
+
\frac{\Pi^{\text{ew}} 4\big|\tau^{\gamma+1\to j } \big|^2 
\text{Im}\big(\rho_+^{0\to\gamma}\big)
\text{Im}\big(\rho_-^{ j +1\to N+1}\big)}
{\big|1- \rho_+^{0\to j } \rho_-^{ j +1\to N+1}\big|^2
\big|1- \rho_+^{0\to\gamma} \rho_-^{\gamma+1\to  j }\big|^2}, 
\qquad  j >\gamma,
\end{split}\label{coeff_seq}\end{equation}
\end{widetext}
whereas, in accordance with Eq.~\eqref{equation_X},
\begin{equation}
\hat{\mathcal{T}}^{N+1}_{\gamma}=\sum_ \ell  X^{\gamma, \ell }=0.
\label{equation_X_2}
\end{equation}
We observe that these coefficients satisfy the symmetry property
\begin{equation}
\hat{\mathcal{T}}^{ j }_{\gamma}=\hat{\mathcal{T}}^{\gamma}_{ j },\qquad j ,\gamma=0,\dots,N.
\label{symmetry_hat_T}
\end{equation}

Furthermore, the radiative heat transfer problem can be equivalently formulated in terms of the coefficients $\hat{\mathcal{T}}^{ j }_{\gamma}$.
According to the definition \eqref{transmission_coefficients_1}, the energy flux \eqref{energy_flux_2} can be rewritten as
\begin{equation}
\Phi_{\gamma}=\int_0^\infty\frac{\dif\omega}{2\pi}\int_0^\infty\frac{\dif  k }{2\pi}\,  k \sum_{p}\sum_{ j =0}^N \hbar\omega n_{ j , j +1}\hat{\mathcal{T}}^{ j }_{\gamma}.
\label{energy_flux_II}
\end{equation}
Moreover, taking into account that the net energy flux on body $ j =1,\dots,N$ is given by $\Phi^{ j }=\Phi_{ j -1}-\Phi_{ j }$, this flux becomes
\begin{equation}
\Phi^{ j }=\int_0^\infty\frac{\dif\omega}{2\pi}\int_0^\infty\frac{\dif  k }{2\pi}\,  k 
\sum_{p}\sum_{ \ell =0}^N \hbar\omega n_{ \ell , \ell +1}
\left(\hat{\mathcal{T}}^{ \ell }_{ j -1}-\hat{\mathcal{T}}^{ \ell }_{ j }\right).
\label{net_energy_flux_II}
\end{equation}
Thus, in view of Eqs.~\eqref{energy_flux_II} and \eqref{net_energy_flux_II}, we see that this formulation is particularly useful when a sequence of consecutive bodies in the system are thermalized at the same temperature; in this case the functions $n_{ \ell , \ell +1}$ cancel out and the corresponding terms do not contribute to the fluxes.

To establish the relation between $\mathcal{T}^{ \ell , j }$ and $\hat{\mathcal{T}}^ j _\gamma$, we first can write
\begin{equation}\begin{split}
X^{\gamma,0} &= \hat{\mathcal{T}}^{0}_{\gamma},\\
X^{\gamma, j } &= \hat{\mathcal{T}}^{ j }_{\gamma} -\hat{\mathcal{T}}^{ j -1}_{\gamma},\qquad  j =1,\dots,N,\\
X^{\gamma,N+1} &= -\hat{\mathcal{T}}^{N}_{\gamma},
\end{split}\label{rel_X_T}\end{equation}
where the first two relations follow directly from the definition \eqref{transmission_coefficients_1}, and the last one is obtained using the fact that $X^{\gamma,N+1} = \hat{\mathcal{T}}^{N+1}_{\gamma} -\hat{\mathcal{T}}^{N}_{\gamma}$ and Eq.~\eqref{equation_X_2}.
Thus, replacing Eq.~\eqref{rel_X_T} in the definition of the energy transmission coefficients given by Eq.~\eqref{trans_coeff} leads to the desired relation:
\begin{equation}\begin{split}
\mathcal{T}^{0, j } &= \hat{\mathcal{T}}^{0}_{ j -1} - \hat{\mathcal{T}}^{0}_{ j },\\
\mathcal{T}^{ \ell , j } &= \hat{\mathcal{T}}^{ \ell }_{ j -1} -\hat{\mathcal{T}}^{ \ell -1}_{ j -1} - \hat{\mathcal{T}}^{ \ell }_{ j } +\hat{\mathcal{T}}^{ \ell -1}_{ j },\\
\mathcal{T}^{N+1, j } &= -\hat{\mathcal{T}}^{N}_{ j -1} + \hat{\mathcal{T}}^{N}_{ j },
\end{split}\label{rel_trans_coeff}\end{equation}
where in the these expressions $ j , \ell =1,\dots,N$.

In addition, we highlight that $X^{\gamma, j }=\hat{\mathcal{T}}^{ j }_{\gamma} -\hat{\mathcal{T}}^{ j -1}_{\gamma}$ vanish if body $ j $ is removed from the system, which can be achieved by letting $\rho_\phi^{ j }\to0$ and $\tau^{ j }\to1$. Accordingly and in view of Eq.~\eqref{rel_trans_coeff}, the associated transmission coefficient $\mathcal{T}^{ \ell , j }$ also vanishes under these conditions, as expected, since this coefficient represents the energy exchange channel between bodies $ \ell $ and $ j $.
Moreover, taking  into account the property \eqref{symmetry_hat_T}, from Eq.~\eqref{rel_trans_coeff} one deduces that the energy transmission coefficients satisfy the reciprocity relation $\mathcal{T}^{ \ell , j }=\mathcal{T}^{ j , \ell }$ for $ j , \ell =1,\dots,N$.

Finally, we note that the contribution of the environmental fields to the coefficients $\hat{\mathcal{T}}^{ j }_{\gamma}$ has to be evaluated by means of Eqs.~\eqref{reflection_transmission_boundary} separately for the cases $j,\gamma=0,N$. In the remaining coefficients $\hat{\mathcal{T}}^{ j }_{\gamma}$, those for which $j,\gamma=1,\dots,N-1$, the contribution of the environmental fields can be straightforwardly evaluated since, for instance, $\rho_+^{0\to j}=\rho_+^{1\to j}$ and $\rho_-^{j+1\to N+1}=\rho_-^{j+1\to N}$. Once this is done, the many-body scattering coefficients have to be expressed using Eqs.~\eqref{hat_coefficients} to remove the dependence of the reflection coefficients on the positions $z_j$ and make explicit the dependence on the separation distances $d_j$ (the whole system is invariant under translations along the $z$ axis).

\section{Momentum transfer coefficients}\label{sec:casimir_coefficients}

In this section we determine the momentum transfer coefficients $\mathcal{W}^{ \ell , j }$ introduced in Eq.~\eqref{W}.
The procedure we adopt here is close to that we followed in Sec.~\ref{sec:transmission_coefficients} to obtain the transmission coefficients $\mathcal{T}^{ \ell , j }$.

In analogy with Eq.~\eqref{transmission_coefficients_1}, we now introduce the set of coefficients $\hat{\mathcal{W}}^{ j }_{\gamma}=\hat{\mathcal{W}}^{ j }_{\gamma}(\omega, k ,p)$ defined as
\begin{equation}
\hat{\mathcal{W}}^{ j }_{\gamma}\equiv \sum_{ \ell =0}^ j  Y^{\gamma, \ell }-\frac{1}{2}\sum_{ \ell } Y^{\gamma, \ell },
\label{transmission_coefficients_2}
\end{equation}
which, in particular, leads to
\begin{equation}
\hat{\mathcal{W}}^{N+1}_{\gamma}=\frac{1}{2}\sum_{ \ell } Y^{\gamma, \ell }.\label{W_N+1}
\end{equation}
Moreover, in Appendix~\ref{derivation_eq_X} we show that the set of coefficients $\hat{\mathcal{W}}^{ j }_{\gamma}$ take the form ($ j ,\gamma=0,\dots,N$)
\begin{widetext}
\begin{equation}\begin{split}
&\hat{\mathcal{W}}^{ j }_\gamma
= \frac{\Pi^{\text{pw}} \big|\tau^{ j +1\to\gamma}\big|^2 \big(1-\big|\rho_+^{0\to j }\big|^2\big)
\big(1 + \big|\rho_-^{\gamma+1\to N+1}\big|^2\big)}
{\big|1 - \rho_+^{0\to j } \rho_-^{ j +1\to \gamma}\big|^2
\big|1- \rho_+^{0\to\gamma} \rho_-^{\gamma+1\to N+1}\big|^2} 
+
\frac{\Pi^{\text{ew}} 4i\big|\tau^{ j +1\to\gamma}\big|^2 \text{Im}\big(\rho_+^{0\to j }\big)
\text{Re}\big(\rho_-^{\gamma+1\to N+1} \big)}
{\big|1 - \rho_+^{0\to j } \rho_-^{ j +1\to \gamma}\big|^2
\big|1- \rho_+^{0\to\gamma} \rho_-^{\gamma+1\to N+1}\big|^2} -\hat{\mathcal{W}}^{N+1}_{\gamma},\,j <\gamma,\\
&\hat{\mathcal{W}}^{\gamma}_\gamma
= \frac{\Pi^{\text{pw}} \big(\big|\rho_-^{\gamma+1\to N+1}\big|^2 -\big|\rho_+^{0\to\gamma}\big|^2\big)}
{\big|1- \rho_+^{0\to\gamma} \rho_-^{\gamma+1\to N+1}\big|^2}
+ 
\frac{\Pi^{\text{ew}} 2i \text{Im}\big(\rho_+^{0\to\gamma}\rho_-^{\gamma+1\to N+1*} \big)}
{\big|1- \rho_+^{0\to\gamma} \rho_-^{\gamma+1\to N+1}\big|^2}\\
&\hat{\mathcal{W}}^{ j }_\gamma
= -\frac{ \Pi^{\text{pw}} \big|\tau^{\gamma+1\to j }\big|^2 \big(1 + \big|\rho_+^{0\to\gamma}\big|^2\big) \big( 1-\big|\rho_-^{ j +1\to N+1} \big|^2 \big)}
{\big|1- \rho_+^{0\to j } \rho_-^{ j +1\to N+1}\big|^2 \big|1 - \rho_+^{0\to\gamma} \rho_-^{\gamma+1\to  j }\big|^2}
-   
\frac{ \Pi^{\text{ew}} 4i\big|\tau^{\gamma+1\to j }\big|^2 \text{Re}\big(\rho_+^{0\to\gamma}\big) \text{Im} \big(\rho_-^{ j +1\to N+1}\big)}
{\big|1- \rho_+^{0\to j } \rho_-^{ j +1\to N+1}\big|^2 \big|1 - \rho_+^{0\to\gamma} \rho_-^{\gamma+1\to  j }\big|^2} +\hat{\mathcal{W}}^{N+1}_{\gamma},\,j >\gamma, \\
&\hat{\mathcal{W}}^{N+1}_{\gamma}
= \frac{\Pi^{\text{pw}} \big(1-\big|\rho_+^{0\to\gamma} \rho_-^{\gamma+1\to N+1}\big|^2\big)}
{\big|1- \rho_+^{0\to\gamma} \rho_-^{\gamma+1\to N+1}\big|^2}
+  
\frac{\Pi^{\text{ew}} 2i \text{Im}\big(\rho_+^{0\to\gamma} \rho_-^{\gamma+1\to N+1}\big)}
{\big|1- \rho_+^{0\to\gamma} \rho_-^{\gamma+1\to N+1}\big|^2} .
\end{split}\label{W_hat_b}\end{equation} 
\end{widetext}

On the other hand, from Eq.~\eqref{transmission_coefficients_2} we can write
\begin{equation}\begin{split}
Y^{\gamma,0} &= \hat{\mathcal{W}}^{0}_{\gamma}+\hat{\mathcal{W}}^{N+1}_{\gamma},\\
Y^{\gamma, j } &= \hat{\mathcal{W}}^{ j }_{\gamma} -\hat{\mathcal{W}}^{ j -1}_{\gamma},\qquad  j =1,\dots,N,\\
Y^{\gamma,N+1} &= -\hat{\mathcal{W}}^{N}_{\gamma}+\hat{\mathcal{W}}^{N+1}_{\gamma}.
\end{split}\end{equation}
Thus, using these relations, from Eq.~\eqref{W} we get
\begin{equation}\begin{split}
\mathcal{W}^{0, j } &= \hat{\mathcal{W}}^{0}_{ j -1} + \hat{\mathcal{W}}^{N+1}_{ j -1} - \hat{\mathcal{W}}^{0}_{ j } - \hat{\mathcal{W}}^{N+1}_{ j } ,\\
\mathcal{W}^{ \ell , j } &= \hat{\mathcal{W}}^{ \ell }_{ j -1} -\hat{\mathcal{W}}^{ \ell -1}_{ j -1} - \hat{\mathcal{W}}^{ \ell }_{ j } +\hat{\mathcal{W}}^{ \ell -1}_{ j }, \\
\mathcal{W}^{N+1, j } &= -\hat{\mathcal{W}}^{N}_{ j -1} + \hat{\mathcal{W}}^{N+1}_{ j -1} + \hat{\mathcal{W}}^{N}_{ j } - \hat{\mathcal{W}}^{N+1}_{ j },
\end{split}\end{equation}
where in these expressions $ j , \ell =1,\dots,N$. This fully determines the coefficients $\mathcal{W}^{ \ell , j }$ in terms of $\hat{\mathcal{W}}^{ j }_{ \gamma }$.

As for the case of energy, the momentum fluxes can be equivalently formulated in terms of the coefficients $\hat{\mathcal{W}}^{ j }_{\gamma}$.
Using the definition \eqref{transmission_coefficients_2}, the momentum flux \eqref{P_gamma_1} can be rewritten as
\begin{equation}\begin{split}
P_{\gamma}&=-\int_0^\infty\frac{\dif\omega}{2\pi}\int_0^\infty\frac{\dif  k }{2\pi}\,  k \sum_{p}\sum_{ j =0}^N \hbar k_z n_{ j , j +1}\hat{\mathcal{W}}^{ j }_{\gamma}\\
&\,-\int_0^\infty\frac{\dif\omega}{2\pi}\int_0^\infty\frac{\dif  k }{2\pi}\,  k \sum_{p}
\hbar k_z \Bigl(\mathcal{N}_0+\mathcal{N}_{N+1}\Bigr)
\hat{\mathcal{W}}^{N+1}_{\gamma}.
\end{split}\label{momentum_flux_II}\end{equation}
Since the net force per unit area acting on body $ j =1,\dots,N$ is given by $P^ j =P_ j -P_{ j -1}$, we get
\begin{equation}\begin{split}
P^{ j }\!&=\!\int_0^\infty\frac{\dif\omega}{2\pi}\int_0^\infty\frac{\dif  k }{2\pi}\,  k 
\sum_{p}\sum_{ \ell =0}^N \hbar k_z n_{ \ell , \ell +1}
\left(\hat{\mathcal{W}}^{ \ell }_{ j -1}-\hat{\mathcal{W}}^{ \ell }_{ j }\right)\\
&\,+\int_0^\infty\frac{\dif\omega}{2\pi}\int_0^\infty\frac{\dif  k }{2\pi}\,  k \sum_p 
\hbar k_z \left(\mathcal{N}_0+\mathcal{N}_{N+1}\right) \\
&\,\times \left(\hat{\mathcal{W}}^{N+1}_{ j -1} -\hat{\mathcal{W}}^{N+1}_{ j }\right).
\end{split}\label{net_momentum_flux_II}\end{equation}
Again, in view of Eq.~\eqref{net_momentum_flux_II}, we see that this formulation is particularly useful when a sequence of consecutive bodies in the system are thermalized at the same temperature. 

We note that taking into account that
\begin{equation}
\sum_ \ell  \mathcal{W}^{ \ell , j } =2\left(\hat{\mathcal{W}}^{N+1}_{ j -1} -\hat{\mathcal{W}}^{N+1}_{ j }\right),
\label{eq_equilibrium}
\end{equation}
at thermal equilibrium Eq.~\eqref{net_momentum_flux_II} becomes $P^j_\mathrm{eq}$ as given by Eq.~\eqref{net_force_equilibrium}. Using Eq.~\eqref{W_hat_b} we immediately deduce that
\begin{equation}
\sum_{j=1}^N\left(\hat{\mathcal{W}}^{N+1}_{ j -1} -\hat{\mathcal{W}}^{N+1}_{ j }\right)=\hat{\mathcal{W}}^{N+1}_0 -\hat{\mathcal{W}}^{N+1}_N=0,
\end{equation}
which proves that the sum of all the forces acting on the bodies vanishes at thermal equilibrium.
In addition, using that $\text{Re}[w/(1-w)]=(1-|w|^2)/(2|1-w|^2)-1/2$ and $\text{Im}[w/(1-w)]=\text{Im}(w)/|1-w|^2$, and the fact that $k_z=|k_z|$ for propagating waves and $k_z=i|k_z|$ for the evanescent ones, from Eqs.~\eqref{W_hat_b} and \eqref{eq_equilibrium}
\begin{equation}\begin{split}
k_z\sum_ \ell  \mathcal{W}^{ \ell , j }&=
4 \text{Re} \left(\frac{ k_z \rho_+^{0\to  j -1} \rho_-^{ j \to N+1} }
{1 -\rho_+^{0\to  j -1} \rho_-^{ j \to N+1} }\right)\\
&\,-4 \text{Re}\left(\frac{k_z \rho_+^{0\to j } \rho_-^{ j +1\to N+1} }
{1 -\rho_+^{0\to j } \rho_-^{ j +1\to N+1}} \right).
\end{split}\end{equation}
Thus, using the above expression, the equilibrium force given by Eq.~(\ref{net_force_equilibrium}) can be more easily calculated by performing a rotation to the imaginary axis~\cite{DzyaloshinskiiAdvPhys61} $\omega\to i\xi$. As a result, the equilibrium force at temperature $T=T_j$ is expressed as a summation over the Matsubara frequencies $\xi_n= 2\pi\kB T n/\hbar$,
\begin{equation}\begin{split}
 &P^ j _\mathrm{eq}= -\frac{\kB T}{\pi}\sum_{n=0}^{\infty}{}'\int_0^\infty\dif  k \,  k  \sum_p \sqrt{\frac{\xi_n^2}{c^2}+ k ^2}\\
 &\,\times\left[\frac{\rho_+^{0\to  j -1} \rho_-^{ j \to N+1}}
{1 -\rho_+^{0\to  j -1} \rho_-^{ j \to N+1} }-\frac{\rho_+^{0\to j } \rho_-^{ j +1\to N+1} }
{1 -\rho_+^{0\to j } \rho_-^{ j +1\to N+1}} \right]_{\omega=i\xi_n},
\end{split}\end{equation}
where the prime in the summation means that the term with $n=0$ has to be multiplied by a factor 1/2, and all the terms in square brackets are evaluated at $\omega=i\xi_n$. We remark that this rotation to the imaginary axis can be performed to deal with the equilibrium contribution to the total pressure, while the thermal nonequilibrium contribution can be computed by integrating over real frequencies.

Analogously to what we said for the coefficients $\hat{\mathcal{T}}^{ j }_{\gamma}$, we emphasize here that the contribution of the environmental fields to $\hat{\mathcal{W}}^{ j }_{\gamma}$ has to be evaluated with Eqs.~\eqref{reflection_transmission_boundary} separately for the cases $j,\gamma=0,N$, and that the many-body scattering coefficients have to be expressed using Eqs.~\eqref{hat_coefficients} to obtain $\hat{\mathcal{W}}^{ j }_{\gamma}$ in terms of the separation distances $d_j$.

\section{Numerical applications}
\label{sec:numerics}

We now illustrate the previous formalism by considering some examples dealing with radiative heat transfer and Casimir-Lifshitz forces in many-body systems. We focus first on temperature configurations leading to local heat transfer equilibrium. Finally, the equilibrium position at which the net force on a given body vanishes is studied by changing the position of another body within the system.
We emphasize here that in the following numerical applications we always consider the steady-state regime.

\subsection{Radiative heat transfer}

As an application of the formulation introduced for heat transfer in many-body systems, we consider now the particular case $N=4$. We are interested in showing how the temperatures corresponding to local heat-transfer equilibrium are modified in four-body systems, as compared with the three-body case. Configurations with bodies at local equilibrium are obtained by letting the temperatures of these bodies reach the particular values for which the net energy flux on them vanishes. Bodies at local equilibrium can thus be considered as passive relays, since a thermostat in contact with each of these bodies will not supply energy to the system under global nonequilibrium conditions. In Appendix~\ref{eq_temp} we discuss a procedure to obtain such equilibrium temperatures.

We analyzed a 4-body system in which two bodies are silicon carbide (SiC) slabs and the other two are made of hexagonal boron nitride (hBN). The permittivity of these polar materials here are described by the Drude-Lorentz model
\begin{equation}
\varepsilon(\omega)=\varepsilon_\infty \frac{\omega^2_L-\omega^2-i\Gamma\omega}{\omega^2_T-\omega^2-i\Gamma\omega}, 
\label{Drude-Lorentz_model}
\end{equation}
with high frequency dielectric constant $\varepsilon_\infty=6.7$, longitudinal optical frequency $\omega_L=1.83\times 10^{14}\,$rad/s, transverse optical frequency $\omega_T=1.49\times 10^{14}\,$rad/s, and damping $\Gamma=8.97\times 10^{11}\,$rad/s for SiC~\cite{Palik}, while for hBN we take~\cite{Eremets95} $\varepsilon_\infty=4.9$, $\omega_L=3.03\times 10^{14}\,$rad/s, $\omega_T=2.57\times 10^{14}\,$rad/s, and $\Gamma=1.0\times 10^{12}\,$rad/s. 

\begin{figure}
\includegraphics[scale=1]{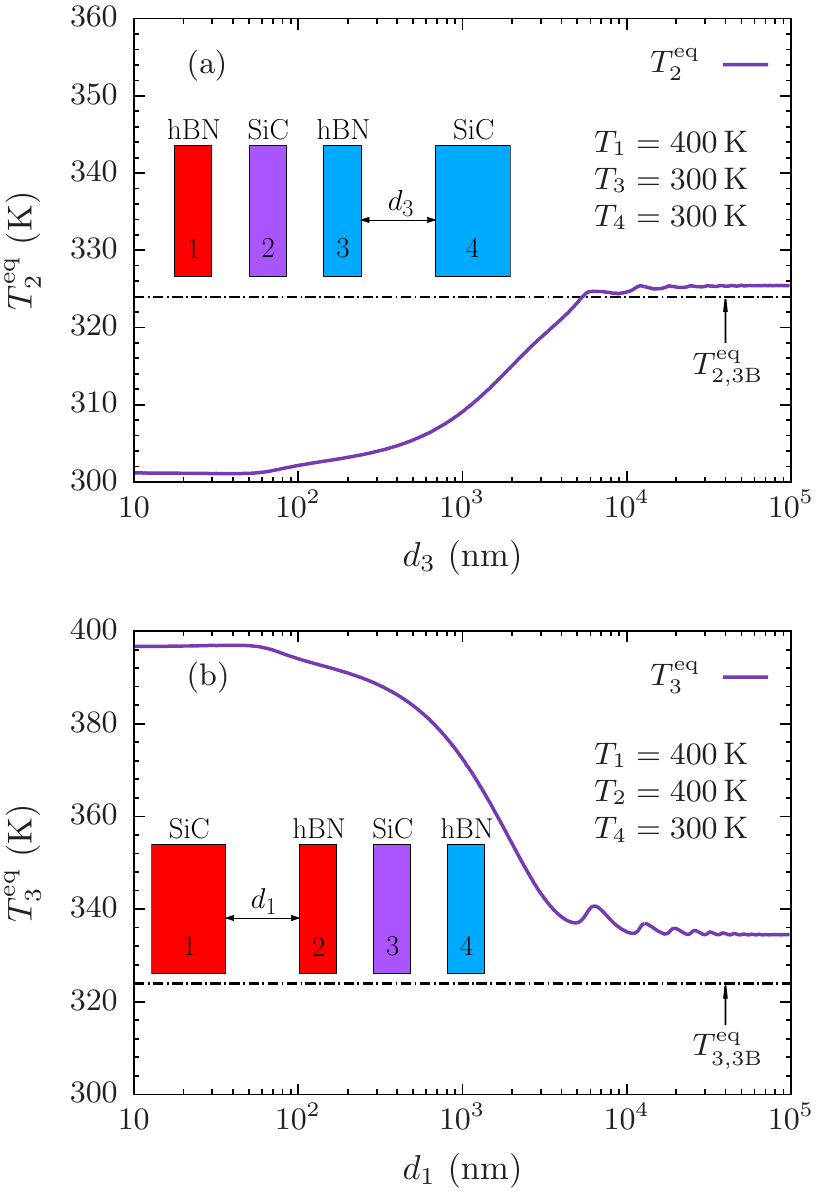} 
\caption{Equilibrium temperatures in four-body systems. (a) Equilibrium temperature of body 2 as a function of the separation distance $d_3$ with fixed $T_1=400\,$K and $T_3=T_4=300\,$K. Bodies 1 and 3 are hBN slabs of width $\delta_1=\delta_3=200\,$nm, whereas bodies 2 and 4 are made of SiC and have widths $\delta_2=200\,$nm and $\delta_4=5\,\mu$m, respectively. The rest of the separation distances are fixed to $d_1=d_2=200\,$nm.
(b) Equilibrium temperature of body 3 as a function of the separation distance $d_1$ with fixed $T_1=T_2=400\,$K and $T_4=300\,$K. Bodies 2 and 4 are hBN slabs of width $\delta_2=\delta_4=200\,$nm, whereas bodies 1 and 3 are made of SiC and have widths $\delta_1=5\,\mu$m and $\delta_3=200\,$nm, respectively. The rest of the separation distances are fixed to $d_2=d_3=200\,$nm.
In (a) and (b), we also show the equilibrium temperature of the intermediate body in a three-body system: $T^\mathrm{eq}_\mathrm{2,3B}$ is obtained by removing body 4 in (a) and $T^\mathrm{eq}_\mathrm{3,3B}$ is obtained by removing body 1 in (b). In both cases the environmental temperatures are fixed to $T_0=T_5=300\,$K.}
\label{fig2}
\end{figure}

Here we consider a setup in which three of these four bodies, corresponding to a hBN-SiC-hBN configuration, are at fixed positions while the position of the remaining SiC slab is varied in two different arrangements. In the first case, bodies 1 and 3 are hBN slabs of width $\delta_1=\delta_3=200\,$nm, and bodies 2 and 4 are SiC slabs of widths $\delta_2=200\,$nm and $\delta_4=5\,\mu$m, respectively. We then fix the separation distances $d_1=200\,$nm (between bodies 1 and 2) and $d_2=200\,$nm (between bodies 2 and 3), and vary the position of body 4 which is specified by the separation distance $d_3$ (between bodies 3 and 4). The temperature of body 1 is set to $T_1=400\,$K, those of bodies 3 and 4 are fixed to $T_3=T_4=300\,$K, and the temperature $T_2$ of body 2 is allowed to reach the value for which the body attains local equilibrium. In Fig.~\ref{fig2}(a), we show the equilibrium temperature $T_2=T^\mathrm{eq}_2$ of body 2 as a function of the separation distance $d_3$. It is observed how body 2 cools down when body 4 is approached at short separation distances. In the second case, bodies 2 and 4 are hBN slabs of width $\delta_2=\delta_4=200\,$nm, whereas bodies 1 and 3 are made of SiC and have widths $\delta_1=5\,\mu$m and $\delta_3=200\,$nm, respectively. In this case, we vary the separation distance $d_1$ and fix the separation distances $d_2=d_3=200\,$nm. The temperatures are taken such that $T_1=T_2=400\,$K, $T_4=300\,$K, and the temperature $T_3$ of body 3 is that for which the body reaches local equilibrium. In Fig.~\ref{fig2}(b), the equilibrium temperature $T_3=T^\mathrm{eq}_3$ of body 3 is shown as a function of the separation distance $d_1$. We observe now that body 3 heats up when body 1 is brought closer to the three-body structure. In Fig.~\ref{fig2}(a) [Fig.~\ref{fig2}(b)] we also show the equilibrium temperature $T^\mathrm{eq}_\mathrm{2,3B}$ [$T^\mathrm{eq}_\mathrm{3,3B}$] of the intermediate body in the three-body system obtained by removing body 4 [body 1]. For symmetry reasons, these two three-body equilibrium temperatures coincide, $T^\mathrm{eq}_\mathrm{2,3B}=T^\mathrm{eq}_\mathrm{3,3B}$. Notice that at large $d_3$ in Fig.~\ref{fig2}(a), the equilibrium temperature $T^\mathrm{eq}_2$ does not converge to $T^\mathrm{eq}_\mathrm{2,3B}$, since also in far field the properties of the fourth body influence the other components of the system. A similar behavior is observed in Fig.~\ref{fig2}(b) at large $d_1$.
In all cases (also in those considered below), the left and right environmental temperatures are fixed to $T_0=T_5=300\,$K.

\begin{figure}
\includegraphics[scale=1]{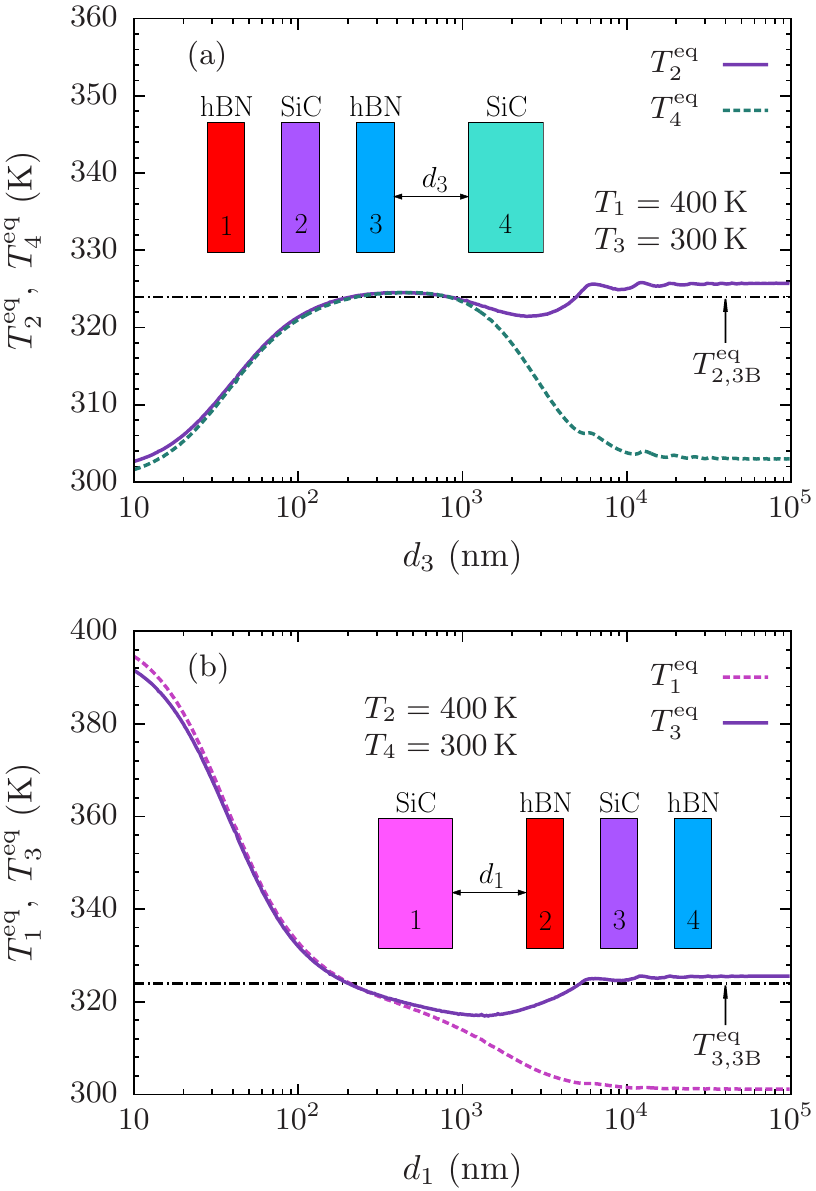} 
\caption{Equilibrium temperatures in four-body systems.  Materials, geometry, and background temperature are the same as in Fig.~\ref{fig2}, but now the temperatures of two bodies are allowed to reach equilibrium conditions. (a) Equilibrium temperatures of bodies 2 and 4 as a function of the separation distance $d_3$ with fixed $T_1=400\,$K and $T_3=300\,$K. (b) Equilibrium temperatures of bodies 1 and 3 as a function of $d_1$ with fixed $T_2=400\,$K and $T_4=300\,$K. For comparison, the equilibrium temperature of the intermediate body in the three-body configuration is also included in the plots.}
\label{fig3}
\end{figure}

We now repeat the same two previous experiences but allowing the SiC slab whose position is varied to attain local equilibrium as well. Hence, now two bodies in the four-body system act as passive relays. For the first case, in Fig.~\ref{fig3}(a) we plot the equilibrium temperatures $T_2=T^\mathrm{eq}_2$ and $T_4=T^\mathrm{eq}_4$ as a function of the separation distance $d_3$ with fixed $T_1=400\,$K and $T_3=300\,$K. In the second case, the equilibrium temperatures $T^\mathrm{eq}_1$ and $T^\mathrm{eq}_3$ are shown in Fig.~\ref{fig3}(b) as a function of $d_1$ with fixed $T_2=400\,$K and $T_4=300\,$K. The equilibrium temperature of the intermediate body in the three-body configuration is also included in the plots.

\begin{figure*}
\includegraphics[scale=0.90]{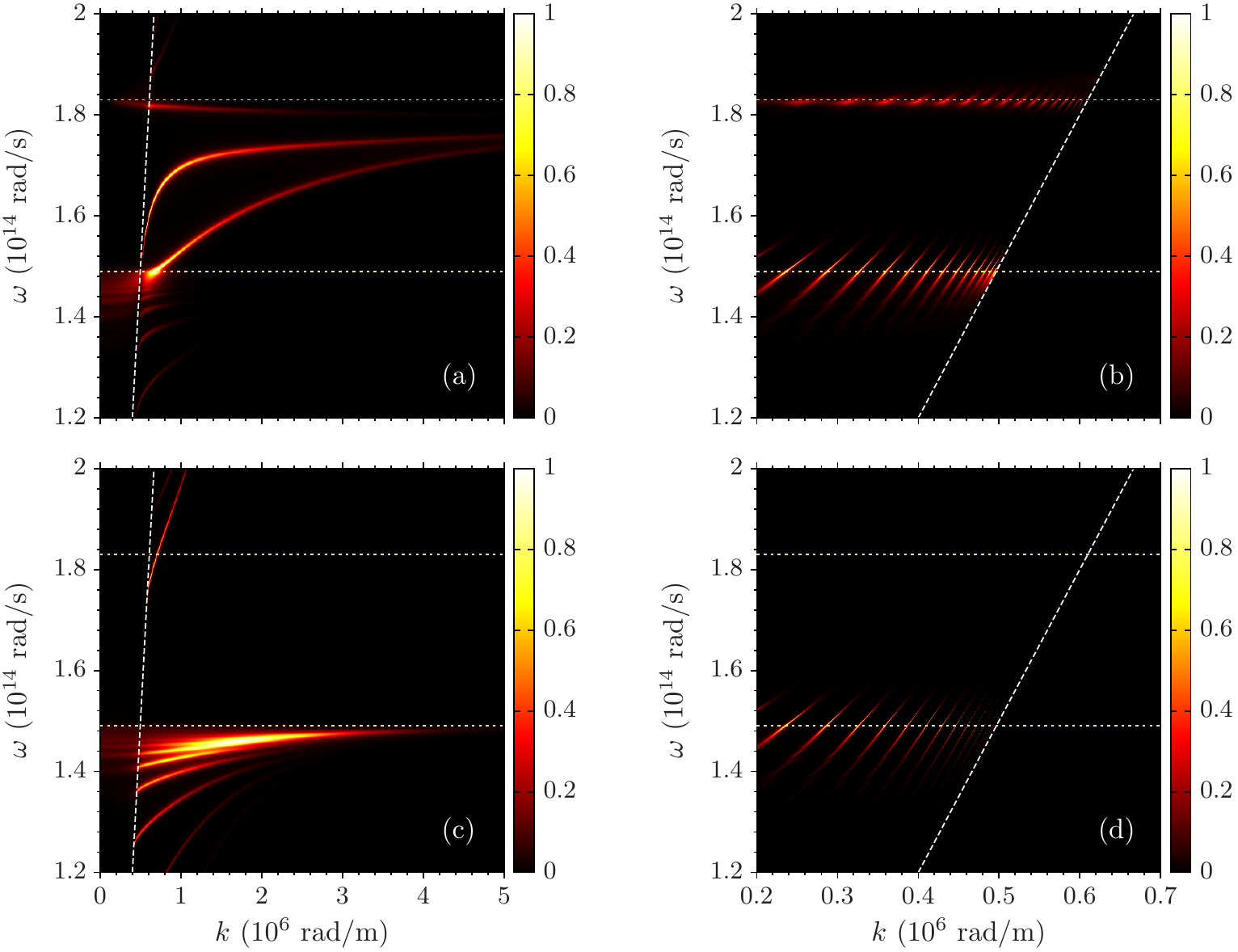} 
\caption{Transmission coefficients $\mathcal{T}^{1,3}$ in the $(k,\omega)$ plane for the four-body configuration corresponding to Fig.~\ref{fig2}(b) and Fig.~\ref{fig3}(b). White dashed lines indicate the light line $\omega=ck$. The upper panel correspond to TM polarization and the lower one to TE polarization. Bodies 1 and 3 are SiC slabs of width $\delta_1=5\,\mu$m and $\delta_3=200\,$nm, respectively, whereas bodies 2 and 4 are made of hBN and have widths $\delta_2=\delta_4=200\,$nm. In (a) and (c) we set $d_1=100\,$nm, while in (b) and (d) we take $d_1=100\,\mu$m. In all cases $d_2=d_3=200\,$nm.
The horizontal lines indicate the longitudinal and transverse optical frequencies $\omega_L=1.83\times 10^{14}\,$rad/s and  $\omega_T=1.49\times 10^{14}\,$rad/s for SiC, respectively.}
\label{fig4}
\end{figure*}

To get insight into the physical mechanism responsible for the heat transfer in the system, we calculate the energy transmission coefficients $\mathcal{T}^{1,3}$ in the ($k,\omega$) plane for two of the previous SiC-hBN-SiC-hBN configurations, corresponding to near and far fields, as shown in Fig.~\ref{fig4}. According to the Landauer formalism, this coefficient can be interpreted as the coupling efficiency of modes between bodies 1 and 3. In Figs.~\ref{fig4}(a) and \ref{fig4}(c), the transmission coefficients are shown in the near-field regime for TM and TE polarizations, respectively. For TM polarization [Fig.~\ref{fig4}(a)], we observe symmetric (low energy) and antisymmetric (high energy) surface phonon-polaritons supported by the SiC samples and an intermediate branch corresponding to a hybridized surface resonance due to the coupling between the two layers. We highlight that these branches are attenuated by the presence of the intermediate hBN slab. For TE polarization [Fig.~\ref{fig4}(c)], the coupling mechanism is radically different, since in this case the system does not support surface waves. However, as shown by the plot of the transmission coefficient, the efficiency of coupling is also important for this polarization. The contribution of TE-polarized waves to the transfer mainly results from frustrated modes which are associated to guided modes in the first SiC layer. Accordingly, the number of branches in Fig.~\ref{fig4}(c) increases with the thickness of body 1. On the other hand, in far-field regime the transfer is due to guided modes in the cavity formed by the first and the third layer. The associated energy transmission coefficients are shown in Figs.~\ref{fig4}(b) and \ref{fig4}(d) for the two polarizations. Finally, we remark that Wien's frequencies $\omega_W(T)=2.82\kB T/\hbar$ indicating the frequencies around which heat exchange occurs are given by $\omega_W(300\,\mathrm{K})=1.11\times10^{14}\,$rad/s and $\omega_W(400\,\mathrm{K})=1.48\times10^{14}\,$rad/s for the temperatures of the thermostated bodies.

\subsection{Casimir-Lifshitz force}

We now consider an application of the formalism to the case of the Casimir-Lifshitz force in a four-body system. Here we restrict ourselves to the case of thermal equilibrium.
Our aim is to show how the equilibrium position of one of the intermediate bodies in the four-body setup is modified by changing the position of one of the external bodies. By equilibrium position we mean the location of the body at which the net Casimir-Lifshitz force on it vanishes. Such an equilibrium position is, however, unstable, because of the purely attractive character of the forces acting on the body.

\begin{figure}
\includegraphics[scale=1]{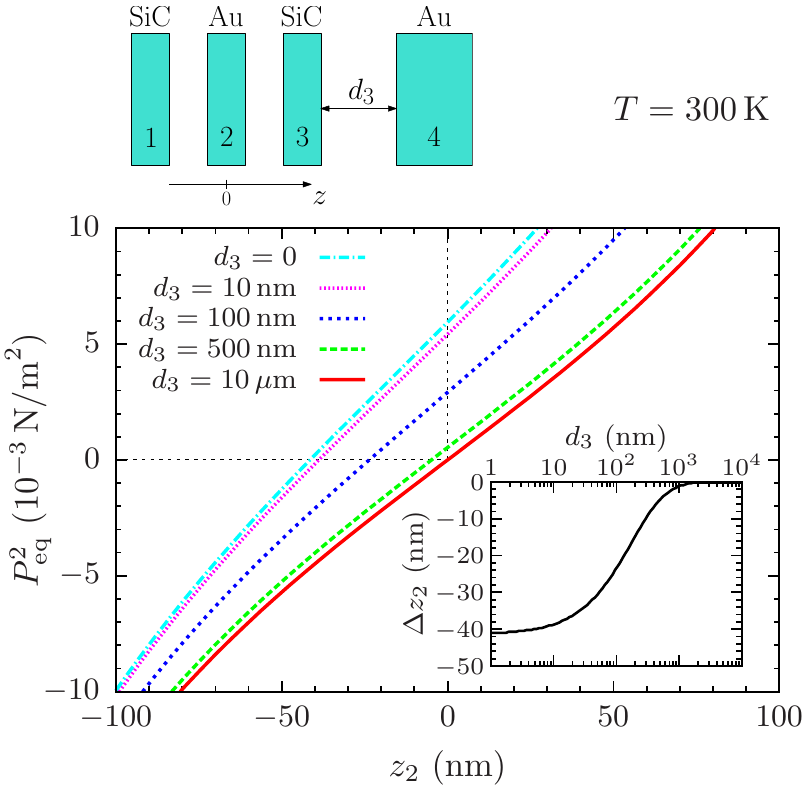} 
\caption{Net pressure $P_\mathrm{eq}^2$ acting on body 2 in a four-body configuration at thermal equilibrium ($T=300\,$K). This pressure is shown as a function of the position of body 2, $z_2$, for several separation distances $d_3$ with fixed $d_1+d_2=1\,\mu$m. Bodies 1 and 3 are SiC slabs of width $\delta_1=\delta_3=100\,$nm, and bodies 2 and 4 are gold slabs of width $\delta_2=100\,$nm and $\delta_4=10\,\mu$m, respectively. The inset shows the variation of the (unstable) equilibrium position of body 2 when $d_3$ is varied. This variation $\Delta z_2$ is measured with respect to the equilibrium position, at $z_2=0$, in the three-body configuration in which body 4 is removed.}
\label{fig5}
\end{figure}

To be more precise, we consider a four-body system in which bodies 1 and 3 are SiC slabs of width $\delta_1=\delta_3=100\,$nm, and bodies 2 and 4 are gold (Au) slabs of width $\delta_2=100\,$nm and $\delta_4=10\,\mu$m, respectively. To describe the permittivity of Au we have used a Drude model
\begin{equation}
\varepsilon(\omega)= 1 - \frac{ \omega^2_P }{ \omega( \omega + i\Gamma ) } ,
\label{Drude_model}
\end{equation}
with plasma frequency $\omega_P=1.37\times 10^{16}\,$rad/s and dissipation rate $\Gamma=5.32\times 10^{13}\,$rad/s. We stress here that our theory can in principle be applied to obtain predictions using any description of optical properties, such as for example the plasma model in the case of metals (see~\cite{KlimchitskayaPRA17} and refs. therein). We then fix the positions of bodies 1 and 3, and compute the net pressure $P_\mathrm{eq}^2$ acting on body 2 as a function of the position $z_2$ of this body for a given location of the fourth slab. We perform this operation for different positions of body 4, which are specified by the separation distance $d_3$ between bodies 3 and 4. Moreover, the bodies are accommodated in such a way that the gaps between bodies 1 and 2 and between 2 and 3 accomplish $d_1+d_2=1\,\mu$m. We measure the position $z_2$ with respect to the center of the cavity formed by bodies 1 and 3. In this way, if the fourth body is absent, by symmetry, the equilibrium position of body 2 is precisely at $z_2=0$, the center of this cavity. The resulting net pressure $P_\mathrm{eq}^2$ is shown in Fig.~\ref{fig5} (the temperature is set to $T=300\,$K). In the inset of Fig.~\ref{fig5}, we also plot the variation of the equilibrium position with respect to the case in which body 4 is not present, $\Delta z_2$, as a function of the separation distance $d_3$. As shown in the plot, in the specific configuration we consider, the presence of the additional fourth gold slab is able to modify the equilibrium position of more than 40\,nm in the best case, i.e. when $d_3=0$, slab 4 being in contact with slab 3.

\section{Conclusions}
\label{sec:conclusions}

In this work we have introduced a general theoretical framework to investigate the radiative heat transfer and the Casimir-Lifshitz force both at and out of thermal equilibrium between bodies separated by vacuum gaps in arbitrary planar many-body systems. A Landauer-like formulation of the heat transfer problem has been carried out and the corresponding energy transmission coefficients have been explicitly expressed in terms of reflection and transmission properties of the different layers. Similar explicit expressions have been derived for coefficients related to momentum transfer.

We have applied this theory to investigate both heat exchanges and interacting forces at thermal equilibrium in systems made of three parallel slabs when a fourth slab is brought close to them. We have shown that near-field interactions can significantly impact heat and momentum exchanges within these systems demonstrating so the potential of many-body interactions to tune these exchanges. In particular, we have numerically proved that varying the position of the fourth slabs allows us to actively tune both the equilibrium temperatures of the system in the case of radiative heat transfer and the mechanical equilibrium position when dealing with Casimir-Lifshitz forces.

Our results show that $N$-body systems are indeed promising candidates for any application where energy- and momentum-exchange manipulation is desired. In particular, the presence of a higher number of degrees of freedom paves the way to a finer control of both effects. Because of the non-additivity of both phenomena, a full $N$-body theory such as the one discussed here is mandatory in order to correctly predict the dependence of heat transfer and force on the system parameters.

\begin{acknowledgments}
I.L. and P.B.-A. thank the French National Center for Scientific Research for supporting their work.
\end{acknowledgments}

\vspace{3mm}

\appendix

\section{Coefficients $\hat{\mathcal{T}}^{ j }_{\gamma}$ and $\hat{\mathcal{W}}^{ j }_{\gamma}$}\label{derivation_eq_X}

Here we first deduce the expressions for the set of coefficients $X^{\gamma, j }$ and $Y^{\gamma, j }$ introduced by Eqs.~\eqref{coefficients_X} and \eqref{coefficients_Y}, respectively. After that we will compute the coefficients  $\hat{\mathcal{T}}^{ j }_{\gamma}$ and $\hat{\mathcal{W}}^{ j }_{\gamma}$ given by Eqs.~\eqref{transmission_coefficients_1} and \eqref{transmission_coefficients_2}, respectively.

For convenience, we now introduce two auxiliary media ``beyond'' the external blackbodies labeled with $ j =-1$ and $ j = N+2$, whose associated reflection and transmission coefficients satisfy
\begin{equation}\begin{split}
\rho_\phi^{-1}&=\rho_\phi^{N+2}=0,\\
\tau^{-1}&=\tau^{N+2}=1.
\end{split}\end{equation}
Introducing these innocuous media can be seen as a trick for fictitiously providing a first-neighbor symmetry around the environmental fields. 
Hence, we see that $\rho_+^{1\to  j }=\rho_+^{0\to  j }=\rho_+^{-1\to  j }$
and $\rho_-^{ j \to N}=\rho_-^{ j \to N+1}=\rho_-^{ j \to N+2}$.
This allows us to rewrite the coefficients $L_{ j -}^{\gamma+}$ and $L_{ j +}^{\gamma-}$ in such a way that from Eqs.~\eqref{L13} and \eqref{L5} we get
\begin{equation}\begin{split}
L_{ j \eta}^{\gamma-}&=\rho_-^{\gamma+1\to N+1} L_{ j \eta}^{\gamma+},\qquad  j \leq\gamma,\\
L_{ j -}^{\gamma+}&= \rho_+^{-1\to  j -1} \tau^{ j } u^{-1\to j -1, j } L_{ j +}^{\gamma+}, \qquad  j \leq\gamma , 
\end{split}\label{L5_bis}\end{equation}
while from Eqs.~\eqref{L13} and \eqref{L6} we have
\begin{equation}\begin{split}
L_{ j \eta}^{\gamma+}&=\rho_+^{0\to\gamma} L_{ j \eta}^{\gamma-},\qquad  j >\gamma,\\
L_{ j +}^{\gamma-}&= \rho_-^{ j +1\to N+2} \tau^{ j } u^{ j , j +1\to N+2} L_{ j -}^{\gamma-}, \quad  j >\gamma.
\end{split}\label{L6_bis}\end{equation}

In view of Eqs.~\eqref{L5_bis} and \eqref{L6_bis}, on the one hand, for $ j \leq\gamma$ the coefficients $L^{\gamma\phi}_{ j \eta}$ are all proportional to $L^{\gamma+}_{ j +}$, the proportionality factor being a function of the many-body scattering coefficients. On the other hand, for $ j >\gamma$ these coefficients are all proportional to $L^{\gamma-}_{ j -}$. This fact allows us to simplify the summations in Eqs.~\eqref{coefficients_X} and \eqref{coefficients_Y}, so that the coefficients $X^{\gamma, j }$ and $Y^{\gamma, j }$ can be more easily obtained separately for $ j \leq\gamma$ and $ j >\gamma$.
In doing so and after replacing the coefficients $\mathcal{K}^{ j \eta\eta'}_{\phi\phi'}$ using Eq.~\eqref{coefficients_K_2}, from Eq.~\eqref{coefficients_X} and for $ j \leq\gamma$ one gets
\begin{widetext}
\begin{equation}\begin{split}
X^{\gamma, j }
&=
\Pi^{\text{pw}}\big|L^{\gamma+}_{ j +} \big|^2 \big(1-\big|\rho_-^{\gamma+1\to N+1}\big|^2\big)
\Big[
1-\big|\rho_+^{ j }\big|^2-\big|\tau^{ j }\big|^2
-\rho_+^{-1\to  j -1*} \tau^{ j *} u^{-1\to j -1, j *}
\big(\rho_+^{ j }\tau^{ j *}+\rho_-^{ j *}\tau^{ j }\big)\\
&\,-
\rho_+^{-1\to  j -1} \tau^{ j } u^{-1\to j -1, j } 
\big(\rho_-^{ j }\tau^{ j *}+\rho_+^{ j *}\tau^{ j }\big)
+
\big|\rho_+^{-1\to  j -1} \tau^{ j } u^{-1\to j -1, j }\big|^2 
\big(1-\big|\rho_-^{ j }\big|^2-\big|\tau^{ j }\big|^2\big) \Big]\\
&\,-
\Pi^{\text{ew}}\big|L^{\gamma+}_{ j +}\big|^2 2i\text{Im}\big(\rho_-^{\gamma+1\to N+1}\big) \Big[
\rho_+^{ j }-\rho_+^{ j *}
+
\rho_+^{-1\to  j -1*} \tau^{ j *} u^{-1\to j -1, j *}
\big(\tau^{ j }-\tau^{ j *}\big)\\
&\,+
\rho_+^{-1\to  j -1} \tau^{ j } u^{-1\to j -1, j }
\big(\tau^{ j }-\tau^{ j *}\big)
+
\big|\rho_+^{-1\to  j -1} \tau^{ j } u^{-1\to j -1, j }\big|^2
\big(\rho_-^{ j }-\rho_-^{ j *}\big)\Big], \qquad  j \leq\gamma.
\end{split}\label{eq_inter_X}\end{equation}
Now, according to the definition of the many-body reflection coefficients, Eq.~\eqref{many-body_sc}, we can write
$\rho_+^{-1\to j }= \rho_+^{ j } + \big(\tau^{ j }\big)^2 u^{-1\to j -1, j } \rho_+^{-1\to  j -1}$. Using this, it is then not difficult to recognize the terms leading to $1-\big|\rho_+^{-1\to j }\big|^2$ in the contribution of propagating waves in square brackets in Eq.~\eqref{eq_inter_X}, and $2i\text{Im}\big(\rho_+^{-1\to j }\big)$ in the terms corresponding to the evanescent sector. Moreover, recalling that $u^{-1\to j -1, j }=\big(1 - \rho_+^{-1\to j -1} \rho_-^ j \big)^{-1}$, the remaining terms can be grouped with the common factor $\big|\tau^{ j } u^{-1\to j -1, j } \big|^2$  in such a way that
\begin{equation}\begin{split}
X^{\gamma, j }
&= \Pi^{\text{pw}} \big|L^{\gamma+}_{ j +}\big|^2 
\big[ 1 -\big|\rho_+^{-1\to j }\big|^2  -\big|\tau^{ j }u^{-1\to j -1, j } \big|^2
\big( 1 -\big|\rho_+^{-1\to  j -1} \big|^2 \big)\big]
\big( 1 -\big|\rho_-^{\gamma+1\to N+1}\big|^2 \big)\\
&\,+ \Pi^{\text{ew}} \big|L^{\gamma+}_{ j +}\big|^2 4
\big[ 
\text{Im}\big(\rho_+^{-1\to j }\big)
-\big|\tau^{ j }u^{-1\to j -1, j }\big|^2
\text{Im}\big(\rho_+^{-1\to  j -1}\big)
\big]
\text{Im}\big(\rho_-^{\gamma+1\to N+1}\big),
\qquad  j \leq\gamma.
\end{split}\label{X_1}\end{equation}
Following a similar procedure for the case $ j >\gamma$, Eq.~\eqref{coefficients_X} can be computed to give
\begin{equation}\begin{split}
X^{\gamma, j }
&= -\Pi^{\text{pw}}\big|L^{\gamma-}_{ j -}\big|^2 \big(1-\big|\rho_+^{0\to\gamma}\big|^2\big)
\big[ 1-\big| \rho_-^{ j \to N+2}\big|^2
-\big|\tau^{ j }u^{ j , j +1\to N+2}\big|^2
\big(1-\big| \rho_-^{ j +1\to N+2}\big|^2 \big)
\big]\\
&\,-\Pi^{\text{ew}}\big|L^{\gamma-}_{ j -}\big|^24\text{Im}\big(\rho_+^{0\to\gamma}\big)
\big[ \text{Im}\big(\rho_-^{ j \to N+2}\big)
-\big|\tau^{ j }u^{ j , j +1\to N+2}\big|^2
\text{Im}\big(\rho_-^{ j +1\to N+2}\big)
\big], 
\qquad  j >\gamma.
\end{split}\label{X_2}\end{equation}

As noted previously, the coefficients $Y^{\gamma, j }$ can be obtained adopting a strategy analogous to the previous one, which for brevity we do not include here.
In summary, using Eqs.~\eqref{L5_bis} and \eqref{L6_bis} to write the coefficients $L^{\gamma\phi}_{ j \eta}$ and \eqref{coefficients_K_2} for the coefficients $\mathcal{K}^{ j \eta\eta'}_{\phi\phi'}$, the summations in Eq.~\eqref{coefficients_Y} can be evaluated yielding
\begin{equation}\begin{split}
Y^{\gamma, j }
&=  \Pi^{\text{pw}}
\big|L^{\gamma+}_{ j +}\big|^2 
\big[ 1-\big|\rho_+^{-1\to j }\big|^2-\big|\tau^{ j }u^{-1\to j -1, j }\big|^2
\big(  1-\big|\rho_+^{-1\to  j -1} \big|^2 \big)\big]
\big(1 + \big|\rho_-^{\gamma+1\to N+1}\big|^2 \big)\\
&\,+ \Pi^{\text{ew}} \big|L^{\gamma+}_{ j +}\big|^2 4i \big[
\text{Im}\big(\rho_+^{-1\to j }\big) -\big|\tau^{ j }u^{-1\to j -1, j }\big|^2
\text{Im}\big(\rho_+^{-1\to  j -1}\big)
\big]
\text{Re}\big(\rho_-^{\gamma+1\to N+1} \big),
\qquad  j \leq\gamma,
\end{split}\label{Y_1}\end{equation}
and
\begin{equation}\begin{split}
Y^{\gamma, j } 
&= \Pi^{\text{pw}} \big|L_{ j -}^{\gamma-} \big|^2
\big( 1 + \big|\rho_+^{0\to\gamma}\big|^2\big)
\big[ 1-\big|\rho_-^{ j \to N+2}\big|^2
-\big|\tau^{ j }u^{ j , j +1\to N+2}\big|^2
\big( 1-\big|\rho_-^{ j +1\to N+2} \big|^2  \big) \big]\\
&\,+ \Pi^{\text{ew}} \big|L_{ j -}^{\gamma-} \big|^2 
4i\text{Re}\big(\rho_+^{0\to\gamma}\big)
\big[ \text{Im}\big(\rho_-^{ j \to N+2}\big) 
-\big|\tau^{ j }u^{ j , j +1\to N+2}\big|^2 
\text{Im} \big(\rho_-^{ j +1\to N+2}\big) \big],
\qquad  j >\gamma.
\end{split}\label{Y_2}\end{equation}

We now focus on the coefficients $\hat{\mathcal{T}}^{ j }_{\gamma}$. The above expressions for $X^{\gamma, j }$, Eqs.~\eqref{X_1} and \eqref{X_2}, can be made more explicit by evaluating them with Eqs.~\eqref{L_plus_plus} and \eqref{L_minus_minus}, which leads to
\begin{equation}\begin{split}
X^{\gamma, j }
&= \Pi^{\text{pw}}
\left[  
\frac{\big|\tau^{ j +1\to\gamma}\big|^2 \big(1-\big|\rho_+^{-1\to j }\big|^2\big)}
{\big|1 - \rho_+^{0\to j } \rho_-^{ j +1\to \gamma}\big|^2}
-\frac{\big|\tau^{ j \to\gamma}\big|^2\big(  1-\big|\rho_+^{-1\to  j -1} \big|^2 \big)}
{\big|1 - \rho_+^{-1\to j -1} \rho_-^{ j \to\gamma}\big|^2}
\right]
\frac{ 1 -\big|\rho_-^{\gamma+1\to N+1}\big|^2 }
{\big|1- \rho_+^{0\to\gamma} \rho_-^{\gamma+1\to N+1}\big|^2}\\
&\,+ \Pi^{\text{ew}} 
\left[
\frac{\big|\tau^{ j +1\to\gamma}\big|^2 \text{Im}\big(\rho_+^{-1\to j }\big)}
{\big|1 - \rho_+^{0\to j } \rho_-^{ j +1\to \gamma}\big|^2}
-\frac{\big|\tau^{ j \to\gamma}\big|^2\text{Im}\big(\rho_+^{-1\to  j -1}\big)}
{\big|1 - \rho_+^{-1\to j -1} \rho_-^{ j \to\gamma}\big|^2}
\right]
\frac{4\text{Im}\big(\rho_-^{\gamma+1\to N+1}\big)}
{\big|1- \rho_+^{0\to\gamma} \rho_-^{\gamma+1\to N+1}\big|^2},
\qquad  j <\gamma,\\
X^{\gamma, \gamma }
&= \Pi^{\text{pw}}
\left[ 
1-\big|\rho_+^{-1\to \gamma }\big|^2-
\frac{\big|\tau^{ \gamma }\big|^2 \big(  1-\big|\rho_+^{-1\to  \gamma -1} \big|^2 \big)}
{\big|1 - \rho_+^{-1\to \gamma -1} \rho_-^{ \gamma }\big|^2}
\right]
\frac{ 1 -\big|\rho_-^{\gamma+1\to N+1}\big|^2 }
{\big|1- \rho_+^{0\to\gamma} \rho_-^{\gamma+1\to N+1}\big|^2}\\
&\,+ \Pi^{\text{ew}} 
\left[
\text{Im}\big(\rho_+^{-1\to \gamma }\big) 
-\frac{\big|\tau^{ \gamma }\big|^2 \text{Im}\big(\rho_+^{-1\to  \gamma -1}\big)}
{\big|1 - \rho_+^{-1\to \gamma -1} \rho_-^{ \gamma }\big|^2}
\right]
\frac{4\text{Im}\big(\rho_-^{\gamma+1\to N+1}\big)}
{\big|1- \rho_+^{0\to\gamma} \rho_-^{\gamma+1\to N+1}\big|^2},\\
X^{\gamma, \gamma+1 }
&= -
\frac{ \Pi^{\text{pw}} \big(1-\big|\rho_+^{0\to\gamma}\big|^2\big) }
{\big|1- \rho_+^{0\to\gamma} \rho_-^{\gamma+1\to N+1}\big|^2}
\left[ 
1-\big|\rho_-^{ \gamma+1 \to N+2}\big|^2
-\frac{\big|\tau^{ \gamma+1 }\big|^2 \big( 1-\big|\rho_-^{ \gamma+2\to N+2} \big|^2 \big)}
{\big|1 - \rho_+^{ \gamma+1 } \rho_-^{ \gamma+2\to N+2}\big|^2}
\right]\\
&\,-
\frac{ \Pi^{\text{ew}} 4\text{Im}\big(\rho_+^{0\to\gamma}\big)}
{\big|1- \rho_+^{0\to\gamma} \rho_-^{\gamma+1\to N+1}\big|^2}
\left[ 
\text{Im}\big(\rho_-^{ \gamma+1 \to N+2}\big) 
-\frac{\big|\tau^{ \gamma+1 }\big|^2 \text{Im} \big(\rho_-^{ \gamma+2\to N+2}\big)}
{\big|1 - \rho_+^{ \gamma+1 } \rho_-^{ \gamma+2\to N+2}\big|^2}
\right],\\
X^{\gamma, j }
&= -
\frac{ \Pi^{\text{pw}} \big( 1-\big|\rho_+^{0\to\gamma}\big|^2 \big) }
{\big|1- \rho_+^{0\to\gamma} \rho_-^{\gamma+1\to N+1}\big|^2}
\left[ 
\frac{\big|\tau^{\gamma+1\to j -1} \big|^2 \big(1-\big|\rho_-^{ j \to N+2}\big|^2\big)}
{\big|1 - \rho_+^{\gamma+1\to j -1} \rho_-^{ j \to N+1}\big|^2}
-\frac{\big|\tau^{\gamma+1\to j }\big|^2 \big( 1-\big|\rho_-^{ j +1\to N+2} \big|^2 \big)}
{\big|1 - \rho_+^{\gamma+1\to j } \rho_-^{ j +1\to N+2}\big|^2}
\right]\\
&\,-
\frac{ \Pi^{\text{ew}} 4\text{Im}\big(\rho_+^{0\to\gamma}\big)}
{\big|1- \rho_+^{0\to\gamma} \rho_-^{\gamma+1\to N+1}\big|^2}
\left[ 
\frac{\big|\tau^{\gamma+1\to j -1} \big|^2 \text{Im}\big(\rho_-^{ j \to N+2}\big)}
{\big|1 - \rho_+^{\gamma+1\to j -1} \rho_-^{ j \to N+1}\big|^2}
-\frac{\big|\tau^{\gamma+1\to j }\big|^2 \text{Im} \big(\rho_-^{ j +1\to N+2}\big)}
{\big|1 - \rho_+^{\gamma+1\to j } \rho_-^{ j +1\to N+2}\big|^2}
\right], 
\qquad  j >\gamma+1, \qquad
\end{split}\label{X_1_2}\end{equation}
where we have used that 
\begin{equation}\begin{split}
&\tau^{ j \to\gamma}=\tau^ j  u^{ j , j +1\to\gamma} \tau^{ j +1\to\gamma},\\
&\frac{u^{0\to j , j +1\to \gamma}u^{-1\to j -1, j }}{u^{ j , j +1\to\gamma} u^{-1\to j -1, j \to\gamma}}=1,
\end{split}\label{rel_12}\end{equation}
for the case $ j <\gamma$, the relations
\begin{equation}\begin{split}
&\tau^{\gamma+1\to j }=\tau^{\gamma+1\to j -1} u^{\gamma+1\to j -1, j } \tau^{ j },\\
&\frac{u^{\gamma+1\to j -1, j \to N+1} u^{ j , j +1\to N+2}}{u^{\gamma+1\to j -1, j } u^{\gamma+1\to j , j +1\to N+2}}=1,
\end{split}\label{rel_34}\end{equation}
for the case $ j >\gamma+1$, and introduced the Fabry-P\'erot denominators using Eq.~\eqref{Fabry-Perot}. We note that the first relations of Eqs.~\eqref{rel_12} and \eqref{rel_34} follow from Eq.~\eqref{many-body_sc}, the definition of the many-body transmission coefficients, while the second ones can be obtained working out Eqs.~\eqref{many-body_sc} and \eqref{Fabry-Perot}. Moreover, by inspection of Eq.~\eqref{X_1_2}, we observe that $X^{\gamma, j }$ are always written as the difference of two terms. Remembering the relation $X^{\gamma, j }=\hat{\mathcal{T}}^{ j }_{\gamma} -\hat{\mathcal{T}}^{ j -1}_{\gamma}$, which is a consequence of the definition \eqref{transmission_coefficients_1} for $ j >0$, allows us to identify $\hat{\mathcal{T}}^{ j }_{\gamma}$ as
\begin{equation}\begin{split}
\hat{\mathcal{T}}^{ j }_{\gamma}
&= 
\frac{\Pi^{\text{pw}} \big|\tau^{ j +1\to\gamma}\big|^2 \big(1-\big|\rho_+^{-1\to j }\big|^2\big) 
\big( 1 -\big|\rho_-^{\gamma+1\to N+1}\big|^2\big)}
{\big|1 - \rho_+^{0\to j } \rho_-^{ j +1\to \gamma}\big|^2 
\big|1- \rho_+^{0\to\gamma} \rho_-^{\gamma+1\to N+1}\big|^2}
+  
\frac{\Pi^{\text{ew}} 4 \big|\tau^{ j +1\to\gamma}\big|^2 \text{Im}\big(\rho_+^{-1\to j }\big) \text{Im}\big(\rho_-^{\gamma+1\to N+1}\big)}
{\big|1 - \rho_+^{0\to j } \rho_-^{ j +1\to \gamma}\big|^2 \big|1- \rho_+^{0\to\gamma} \rho_-^{\gamma+1\to N+1}\big|^2},\qquad  j <\gamma,
\end{split}\label{T_hat_1}\end{equation}
and
\begin{equation}\begin{split}
\hat{\mathcal{T}}^{\gamma}_{\gamma}
&= \Pi^{\text{pw}}
\frac{ \big( 1-\big|\rho_+^{-1\to\gamma}\big|^2 \big) 
\big( 1 -\big|\rho_-^{\gamma+1\to N+1}\big|^2 \big)}
{\big|1- \rho_+^{0\to\gamma} \rho_-^{\gamma+1\to N+1}\big|^2}
+ \Pi^{\text{ew}} 
\frac{4\text{Im}\big(\rho_+^{-1\to\gamma}\big) 
\text{Im}\big(\rho_-^{\gamma+1\to N+1}\big)}
{\big|1- \rho_+^{0\to\gamma} \rho_-^{\gamma+1\to N+1}\big|^2}.
\end{split}\label{T_hat_2}\end{equation}
In fact, the previous identification can be made except for and additive function of $\gamma$ that cancels out when computing the difference $X^{\gamma, j }=\hat{\mathcal{T}}^{ j }_{\gamma} -\hat{\mathcal{T}}^{ j -1}_{\gamma}$. However, since the above expressions lead to $\hat{\mathcal{T}}^{0}_{\gamma}=X^{\gamma,0}$, in agreement with \eqref{transmission_coefficients_1}, such an additive contribution is actually zero.
Analogously, from Eq.~\eqref{X_1_2} we see that
\begin{equation}\begin{split}
\hat{\mathcal{T}}^{ j }_{\gamma}
&= 
\frac{ \Pi^{\text{pw}} \big|\tau^{\gamma+1\to j }\big|^2
\big( 1-\big|\rho_+^{0\to\gamma}\big|^2 \big) 
\big( 1-\big|\rho_-^{ j +1\to N+2} \big|^2 \big)}
{\big|1- \rho_+^{0\to j } \rho_-^{ j +1\to N+2}\big|^2
\big|1 - \rho_+^{0\to\gamma} \rho_-^{\gamma+1\to  j }\big|^2}
+
\frac{ \Pi^{\text{ew}} 4 \big|\tau^{\gamma+1\to j }\big|^2
\text{Im}\big(\rho_+^{0\to\gamma}\big)
\text{Im} \big(\rho_-^{ j +1\to N+2}\big)}
{\big|1- \rho_+^{0\to j } \rho_-^{ j +1\to N+2}\big|^2
\big|1 - \rho_+^{0\to\gamma} \rho_-^{\gamma+1\to  j }\big|^2},\qquad  j >\gamma,
\end{split}\label{T_hat_3}\end{equation}
where we have rewritten the denominators using Eq.~\eqref{Fabry-Perot} and
\begin{equation}
\frac{u^{0\to\gamma,\gamma+1\to N+1} u^{\gamma+1\to j , j +1\to N+2}}{u^{0\to j , j +1\to N+2} u^{0\to\gamma,\gamma+1\to  j } } =1.
\label{rel_5}
\end{equation}
Finally, since $\tau^{N+1}=0$, the coefficient $\tau^{\gamma+1\to j }$ vanishes at $ j =N+1$ and hence, from Eq.~\eqref{T_hat_3} we get $\hat{\mathcal{T}}^{N+1}_{\gamma}=0$. According to the definition of these coefficients, the fact that $\hat{\mathcal{T}}^{N+1}_{\gamma}=\sum_ j  X^{\gamma, j }=0$ shows that Eq.~\eqref{equation_X} indeed holds. Thus, after removing the dependence on the auxiliary media $ j =-1$ and $ j =N+2$, from Eqs.~\eqref{T_hat_1}, \eqref{T_hat_2}, and \eqref{T_hat_3}, we obtain Eqs.~\eqref{coeff_seq} and \eqref{equation_X_2}.
 
Now we turn our attention to the coefficients $\hat{\mathcal{W}}^{ j }_{\gamma}$ given by Eq.~\eqref{transmission_coefficients_2}. 
Evaluating Eqs.~\eqref{Y_1} and \eqref{Y_2} with Eqs.~\eqref{L_plus_plus} and \eqref{L_minus_minus} leads to
\begin{equation}\begin{split}
Y^{\gamma, j }
&=  \Pi^{\text{pw}} 
\left[  
\frac{\big|\tau^{ j +1\to\gamma}\big|^2 \big(1-\big|\rho_+^{-1\to j }\big|^2\big)}
{\big|1 - \rho_+^{0\to j } \rho_-^{ j +1\to \gamma}\big|^2}
-\frac{\big|\tau^{ j \to\gamma}\big|^2\big(  1-\big|\rho_+^{-1\to  j -1} \big|^2 \big)}
{\big|1 - \rho_+^{-1\to j -1} \rho_-^{ j \to\gamma}\big|^2}
\right]
\frac{1 + \big|\rho_-^{\gamma+1\to N+1}\big|^2}
{\big|1- \rho_+^{0\to\gamma} \rho_-^{\gamma+1\to N+1}\big|^2}\\
&\,+ \Pi^{\text{ew}} 
\left[
\frac{\big|\tau^{ j +1\to\gamma}\big|^2 \text{Im}\big(\rho_+^{-1\to j }\big)}
{\big|1 - \rho_+^{0\to j } \rho_-^{ j +1\to \gamma}\big|^2}
-\frac{\big|\tau^{ j \to\gamma}\big|^2\text{Im}\big(\rho_+^{-1\to  j -1}\big)}
{\big|1 - \rho_+^{-1\to j -1} \rho_-^{ j \to\gamma}\big|^2}
\right]
\frac{ 4i\text{Re}\big(\rho_-^{\gamma+1\to N+1} \big)}
{\big|1- \rho_+^{0\to\gamma} \rho_-^{\gamma+1\to N+1}\big|^2},
\qquad  j <\gamma,\\
Y^{\gamma, \gamma }
&=  \Pi^{\text{pw}}
\left[ 
1-\big|\rho_+^{-1\to \gamma }\big|^2-
\frac{\big|\tau^{ \gamma }\big|^2 \big(  1-\big|\rho_+^{-1\to  \gamma -1} \big|^2 \big)}
{\big|1 - \rho_+^{-1\to \gamma -1} \rho_-^{ \gamma }\big|^2}
\right]
\frac{1 + \big|\rho_-^{\gamma+1\to N+1}\big|^2}
{\big|1- \rho_+^{0\to\gamma} \rho_-^{\gamma+1\to N+1}\big|^2}\\
&\,+ \Pi^{\text{ew}} 
\left[
\text{Im}\big(\rho_+^{-1\to \gamma }\big) 
-\frac{\big|\tau^{ \gamma }\big|^2 \text{Im}\big(\rho_+^{-1\to  \gamma -1}\big)}
{\big|1 - \rho_+^{-1\to \gamma -1} \rho_-^{ \gamma }\big|^2}
\right]
\frac{4i \text{Re}\big(\rho_-^{\gamma+1\to N+1} \big)}
{\big|1- \rho_+^{0\to\gamma} \rho_-^{\gamma+1\to N+1}\big|^2},\\
Y^{\gamma, \gamma+1 } 
&= 
\frac{ \Pi^{\text{pw}} \big( 1 + \big|\rho_+^{0\to\gamma}\big|^2 \big)}
{\big|1- \rho_+^{0\to\gamma} \rho_-^{\gamma+1\to N+1}\big|^2}
\left[ 
1-\big|\rho_-^{ \gamma+1 \to N+2}\big|^2
-\frac{\big|\tau^{ \gamma+1 }\big|^2 \big( 1-\big|\rho_-^{ \gamma+2\to N+2} \big|^2 \big)}
{\big|1 - \rho_+^{ \gamma+1 } \rho_-^{ \gamma+2\to N+2}\big|^2}
\right]\\
&\,+  
\frac{\Pi^{\text{ew}} 4i\text{Re}\big(\rho_+^{0\to\gamma}\big)}
{\big|1- \rho_+^{0\to\gamma} \rho_-^{\gamma+1\to N+1}\big|^2}
\left[ 
\text{Im}\big(\rho_-^{ \gamma+1 \to N+2}\big) 
-\frac{\big|\tau^{ \gamma+1 }\big|^2 \text{Im} \big(\rho_-^{ \gamma+2\to N+2}\big)}
{\big|1 - \rho_+^{ \gamma+1 } \rho_-^{ \gamma+2\to N+2}\big|^2}
\right],\\
Y^{\gamma, j } 
&= 
\frac{ \Pi^{\text{pw}} \big( 1 + \big|\rho_+^{0\to\gamma}\big|^2 \big)}
{\big|1- \rho_+^{0\to\gamma} \rho_-^{\gamma+1\to N+1}\big|^2}
\left[ 
\frac{\big|\tau^{\gamma+1\to j -1} \big|^2 \big(1-\big|\rho_-^{ j \to N+2}\big|^2\big)}
{\big|1 - \rho_+^{\gamma+1\to j -1} \rho_-^{ j \to N+1}\big|^2}
-\frac{\big|\tau^{\gamma+1\to j }\big|^2 \big( 1-\big|\rho_-^{ j +1\to N+2} \big|^2 \big)}
{\big|1 - \rho_+^{\gamma+1\to j } \rho_-^{ j +1\to N+2}\big|^2}
\right]\\
&\,+   
\frac{ \Pi^{\text{ew}} 4i\text{Re}\big(\rho_+^{0\to\gamma}\big)}
{\big|1- \rho_+^{0\to\gamma} \rho_-^{\gamma+1\to N+1}\big|^2}
\left[ 
\frac{\big|\tau^{\gamma+1\to j -1} \big|^2 \text{Im}\big(\rho_-^{ j \to N+2}\big)}
{\big|1 - \rho_+^{\gamma+1\to j -1} \rho_-^{ j \to N+1}\big|^2}
-\frac{\big|\tau^{\gamma+1\to j }\big|^2 \text{Im} \big(\rho_-^{ j +1\to N+2}\big)}
{\big|1 - \rho_+^{\gamma+1\to j } \rho_-^{ j +1\to N+2}\big|^2}
\right],
\qquad  j >\gamma+1,
\end{split}\label{Y_abcd}\end{equation}
where, as before, we have used Eq.~\eqref{rel_12} for $ j <\gamma$, Eq.~\eqref{rel_34} for $ j >\gamma+1$, and introduced the Fabry-P\'erot denominators employing Eq.~\eqref{Fabry-Perot}.
Since, in accordance with Eq.~\eqref{transmission_coefficients_2}, we can write $Y^{\gamma, j }= \hat{\mathcal{W}}^{ j }_\gamma- \hat{\mathcal{W}}^{ j -1}_\gamma$ for $ j >0$, by inspection of Eq.~\eqref{Y_abcd} we identify 
\begin{equation}\begin{split}
\hat{\mathcal{W}}^{ j }_\gamma
&=   
\frac{\Pi^{\text{pw}} \big|\tau^{ j +1\to\gamma}\big|^2 \big(1-\big|\rho_+^{-1\to j }\big|^2\big)
\big(1 + \big|\rho_-^{\gamma+1\to N+1}\big|^2\big)}
{\big|1 - \rho_+^{0\to j } \rho_-^{ j +1\to \gamma}\big|^2
\big|1- \rho_+^{0\to\gamma} \rho_-^{\gamma+1\to N+1}\big|^2}\\
&\,+  
\frac{\Pi^{\text{ew}} 4i\big|\tau^{ j +1\to\gamma}\big|^2 \text{Im}\big(\rho_+^{-1\to j }\big)
\text{Re}\big(\rho_-^{\gamma+1\to N+1} \big)}
{\big|1 - \rho_+^{0\to j } \rho_-^{ j +1\to \gamma}\big|^2
\big|1- \rho_+^{0\to\gamma} \rho_-^{\gamma+1\to N+1}\big|^2} +f(\gamma), \qquad  j <\gamma,
\end{split}\label{W_hat_1}\end{equation}
and
\begin{equation}\begin{split}
\hat{\mathcal{W}}^{\gamma}_\gamma
&=  \Pi^{\text{pw}}
\frac{\big(1-\big|\rho_+^{-1\to\gamma}\big|^2\big) \big(1 + \big|\rho_-^{\gamma+1\to N+1}\big|^2\big)}
{\big|1- \rho_+^{0\to\gamma} \rho_-^{\gamma+1\to N+1}\big|^2}
+ \Pi^{\text{ew}} 
\frac{4i \text{Im}\big(\rho_+^{-1\to\gamma}\big) \text{Re}\big(\rho_-^{\gamma+1\to N+1} \big)}
{\big|1- \rho_+^{0\to\gamma} \rho_-^{\gamma+1\to N+1}\big|^2} + f(\gamma),
\end{split}\label{W_hat_2}\end{equation}
where $f(\gamma)$ is to be found. Due to the fact that the above expressions for $ j =0$ reduce to $\hat{\mathcal{W}}^{0}_\gamma=Y^{\gamma,0}+f(\gamma)$, in accordance with Eq.~\eqref{transmission_coefficients_2} we see that $f(\gamma)=-\frac{1}{2}\sum_ j  Y^{\gamma, j }$.
Furthermore, the particular case $ j =\gamma+1$ can be computed as $\hat{\mathcal{W}}^{\gamma+1}_\gamma=Y^{\gamma,\gamma+1}+\hat{\mathcal{W}}^{\gamma}_\gamma$, yielding
\begin{equation}\begin{split}
\hat{\mathcal{W}}^{\gamma+1}_\gamma 
&=- 
\frac{\Pi^{\text{pw}} \big|\tau^{\gamma+1}\big|^2 
\big(1 + \big|\rho_+^{0\to\gamma}\big|^2\big)
\big( 1-\big|\rho_-^{\gamma+2\to N+2} \big|^2 \big)}
{\big|1- \rho_+^{0\to\gamma} \rho_-^{\gamma+1\to N+1}\big|^2
\big|1 - \rho_+^{\gamma+1} \rho_-^{\gamma+2\to N+2}\big|^2}\\
&\,- 
\frac{\Pi^{\text{ew}}  4i \big|\tau^{\gamma+1}\big|^2 
\text{Re}\big(\rho_+^{0\to\gamma}\big)
\text{Im}\big(\rho_-^{\gamma+2\to N+2}\big)}
{\big|1- \rho_+^{0\to\gamma} \rho_-^{\gamma+1\to N+1}\big|^2 
\big|1 - \rho_+^{\gamma+1} \rho_-^{\gamma+2\to N+2}\big|^2}
-\frac{1}{2}\sum_ j  Y^{\gamma, j } + g(\gamma),
\end{split}\end{equation}
where
\begin{equation}\begin{split}
g(\gamma)
&= \Pi^{\text{pw}}
\frac{2\big(1-\big|\rho_+^{0\to\gamma} \rho_-^{\gamma+1\to N+1}\big|^2\big)}
{\big|1- \rho_+^{0\to\gamma} \rho_-^{\gamma+1\to N+1}\big|^2}
+ \Pi^{\text{ew}} 
\frac{4i \text{Im}\big(\rho_+^{0\to\gamma} \rho_-^{\gamma+1\to N+1}\big)}
{\big|1- \rho_+^{0\to\gamma} \rho_-^{\gamma+1\to N+1}\big|^2} .
\end{split}\end{equation}
By comparison of the previous expression for $\hat{\mathcal{W}}^{\gamma+1}_\gamma$ with $\hat{\mathcal{W}}^{ j }_\gamma = Y^{\gamma, j }+ \hat{\mathcal{W}}^{ j -1}_\gamma$ for $ j >\gamma+1$, from Eq.~\eqref{Y_abcd} we infer that
\begin{equation}\begin{split}
\hat{\mathcal{W}}^{ j }_\gamma
&=- 
\frac{ \Pi^{\text{pw}} \big|\tau^{\gamma+1\to j }\big|^2 \big(1 + \big|\rho_+^{0\to\gamma}\big|^2\big) \big( 1-\big|\rho_-^{ j +1\to N+2} \big|^2 \big)}
{\big|1- \rho_+^{0\to j } \rho_-^{ j +1\to N+2}\big|^2 \big|1 - \rho_+^{0\to\gamma} \rho_-^{\gamma+1\to  j }\big|^2}\\
&\,-   
\frac{ \Pi^{\text{ew}} 4i\big|\tau^{\gamma+1\to j }\big|^2 \text{Re}\big(\rho_+^{0\to\gamma}\big) \text{Im} \big(\rho_-^{ j +1\to N+2}\big)}
{\big|1- \rho_+^{0\to j } \rho_-^{ j +1\to N+2}\big|^2 \big|1 - \rho_+^{0\to\gamma} \rho_-^{\gamma+1\to  j }\big|^2}
-\frac{1}{2}\sum_ j  Y^{\gamma, j } + g(\gamma),\qquad  j >\gamma,
\end{split}\label{W_hat_3}\end{equation}
where we have arranged the denominators in the two first terms using Eq.~\eqref{rel_5}. Moreover, evaluating Eq.~\eqref{W_hat_3} at $ j =N+1$ yields $\hat{\mathcal{W}}^{N+1}_\gamma= -\frac{1}{2}\sum_ j  Y^{\gamma, j } + g(\gamma)$, since the coefficient $\tau^{\gamma+1\to  j }$ vanishes when $ j $ corresponds to an environmental field. Thus, this result and Eq.~\eqref{W_N+1} readily lead to the identification
\begin{equation}
\hat{\mathcal{W}}^{N+1}_\gamma=\frac{1}{2}\sum_ j  Y^{\gamma, j }=\frac{1}{2}g(\gamma).\label{g_app}
\end{equation}

Finally, removing the dependence on the auxiliary media $ j =-1$ and $ j =N+2$, from Eqs.~\eqref{W_hat_1}, \eqref{W_hat_2}, \eqref{W_hat_3}, and \eqref{g_app}, we obtain Eq.~\eqref{W_hat_b}.
\end{widetext}

\section{Temperature configurations of local heat transfer equilibrium}
\label{eq_temp}

Here we present a method to find temperature configurations in the $N$-body system for which a given number of bodies within the system are allowed to reach local heat transfer equilibrium. Such configurations are defined by the fact that the net energy flux on these bodies vanishes.

We start by noting that, using Eqs.~\eqref{equation_X} and \eqref{trans_coeff}, the net energy flux \eqref{total_energy_flux_2} can be rewritten as
\begin{equation}
\Phi^{ j }=\int_0^\infty\frac{\dif\omega}{2\pi}\hbar\omega\sum_ \ell   n_{ \ell }(\omega) Q^{ \ell , j }(\omega) ,
\label{flux_Q}
\end{equation}
where
\begin{equation}
Q^{ \ell , j }(\omega)= \int_0^\infty\frac{\dif  k }{2\pi}\,  k \sum_{p} \mathcal{T}^{ \ell , j }(\omega, k ,p).
\end{equation}
Assuming that the transmission coefficients do not depend on temperature, expression \eqref{flux_Q} is convenient for our purpose because each term of the sum over $ \ell $ depends only on one of the temperatures $T_ \ell $.

In order to determine the local equilibrium configurations, consider that $m$ bodies are not thermalized with a bath, so that their temperatures can evolve to reach a configuration of local heat transfer equilibrium, while $N-m$ bodies have fixed temperature. The temperatures of the environments are assumed fixed as well.
To proceed, we introduce the vectors 
\begin{equation}\begin{split}
\vect{x}&=\left(T_{ \ell _1},T_{ \ell _2},\dots,T_{ \ell _m}\right)\in\mathbb{R}^m, \\
\vect{f}&=\left(\Phi^{ \ell _1},\Phi^{ \ell _2},\dots,\Phi^{ \ell _m}\right)\in\mathbb{R}^m, \vphantom{\int}
\end{split}\end{equation}
where $ \ell _1, \ell _2,\dots, \ell _m$ are the bodies that are not thermalized with a bath (not necessarily consecutive). 
Thus, the local equilibrium condition is given by $\vect{x}_e$ satisfying $\vect{f}(\vect{x}_e)=\vect{0}$, which corresponds to the solution of a nonlinear system of equations. It is possible, however, to obtain such a configuration solving linear systems of equations by means of an iterative procedure, as briefly discussed below.

The linear expansion of $\vect{f}(\vect{x})$ around the point $\vect{x}_0$ is given by
\begin{equation}
\vect{f}(\vect{x})\approx\vect{f}(\vect{x}_0)+J(\vect{x}_0)\Delta\vect{x},  
\end{equation}
where $\Delta\vect{x}=\vect{x}-\vect{x}_0$ and $J(\vect{x})=D\vect{f}(\vect{x})$ is the associated Jacobian matrix.
Since from \eqref{flux_Q} we can write
\begin{equation}
\frac{\partial\Phi^ j }{\partial T_ \ell }= \int_0^\infty\frac{\dif\omega}{2\pi} \hbar\omega \frac{\partial n_ \ell (\omega)}{\partial T_ \ell } Q^{ \ell , j }(\omega),
\label{partial_Phi_temperature}
\end{equation}
the components $J_{ij}$ of the Jacobian take the form ($i,j=1,\dots,m$)
\begin{equation}
J_{ij}= 
\int_0^\infty\frac{\dif\omega}{2\pi} \hbar\omega \frac{\partial n_{ \ell _j}(\omega)}{\partial T_{ \ell _j}} Q^{ \ell _j, \ell _i}(\omega).
\end{equation}
Therefore, solving for $\Delta\vect{x}$ the linear system of equations $-\vect{f}(\vect{x}_0)=J(\vect{x}_0)\Delta\vect{x}$,
the equilibrium temperatures are obtained as $\vect{x}_e =\Delta\vect{x}+\vect{x}_0$.
Starting from a given point $\vect{x}_0$, the process can be iterated using $\vect{x}_e$ as the new initial value.

\end{document}